\documentclass[preprint2]{aastex}

\def\deg{\hbox{$^\circ$}}
\def\sun{\hbox{$\odot$}}
\def\farcm{\hbox{$.\mkern-4mu^\prime$}}
\def\farcs{\hbox{$.\!\!^{\prime\prime}$}}

\defcitealias{smi05}{Paper I}

\shorttitle{The Tidal Arm of Arp 107}
\shortauthors{Lapham et al.}

\begin{document}

\title{UV/Optical/IR Color Sequences Along the Tidal Ring/Arm of Arp 107}

\author{Ryen C. Lapham\altaffilmark{1}
and Beverly J. Smith}
\affil{Department of Physics and Astronomy, East Tennessee State University, Johnson City, TN 37614; rlapham@nmt.edu, smithbj@etsu.edu}
\altaffiltext{1}{Now at the Department of Physics, New Mexico Tech, Socorro NM  87801}

\author{and Curtis Struck}
\affil{Department of Physics and Astronomy, Iowa State University, 
Ames, IA 50011; curt@iastate.edu}

\begin{abstract}

We construct UV/optical/IR spectral energy distributions for 29 star forming regions
in the interacting galaxy Arp 107,
using GALEX UV, Sloan Digitized Sky Survey optical, and Spitzer infrared images.
In an earlier study utilizing only the Spitzer data, we found a sequence in 
the mid-infrared colors of star-forming knots along the strong tidal arm in this system.
In the current study, we 
find sequences in the UV/optical colors along the tidal arm that mirror
those in the mid-infrared, with blue UV/optical colors found for regions
that are red in the mid-infrared, and vice versa.
With single-burst stellar population synthesis models, we 
find a sequence in the average stellar age along this arm,
with younger stars preferentially located further out in the arm.
Models that allow two populations of different ages and dust
attenuations suggest
that there may be both
a young component 
and an older population present in these regions.
Thus
the observed color sequences may be better interpreted 
as a sequence
in the relative proportion of young and old stars along the arm, 
with a larger
fraction of young stars near the end.
Comparison with star forming regions in other interacting galaxies
shows that the Arp 107 regions are relatively quiescent, with less intense
star formation than in many other systems.

\end{abstract}

\keywords{galaxies: individual (Arp 107)--galaxies: interactions--galaxies: starburst}

\section{Introduction}

Close encounters and collisions between galaxies can dramatically alter
their morphologies.  Depending upon the parameters
of the interaction, such encounters can produce a wide variety of 
structures including
long tails, bridges, and collisional rings
(see \citealp{str99} and \citealp{duc11} for reviews).
Interactions can also trigger the formation of new stars,
enhancing the overall rates of star formation
\citep{bushouse87, kennicutt87, bushouse88, barton00, 
barton03, smi07}.
The star formation in interacting
galaxies is often quite centrally-concentrated 
(e.g., \citealp{lonsdale84, kee85, 
smi07}),  
however, tidal features frequently have on-going star formation as well
(e.g., \citealp{schweizer78, schombert90, smi10}).
In some systems, massive concentrations of stars and gas (`tidal dwarf
galaxies')
are present near the ends of tails 
\citep{duc97, duc00, smi10}.
Along tidal features 
regularly-spaced knots of star formation (`beads on a string') 
are sometimes seen
\citep{hancock07, smi10}, likely caused by 
local gravitational instabilities (\citealp{elmegreen96}).
Very luminous star forming regions are also sometimes found at the 
base of tidal features,
the so-called 
`hinge clumps' \citep{hancock09, smi10}.
These may be produced by intersecting caustics, where 
a caustic is a narrow pile-up zone caused by orbit-crossing 
in interacting
systems \citep{struck12}.
Gas flows between galaxies in interacting pairs
can also sometimes
trigger star formation in polar rings or
unusual tail-like structures
\citep{bournaud03, struck03, smi08}.

In the current paper, we investigate star formation processes
within the interacting galaxy
pair Arp 107.
Arp 107 is one of about three dozen
nearby pre-merger interacting
galaxies in the `Spirals, Bridges, and Tails' (SB\&T) sample, for which we 
have acquired 
Spitzer IR
and Galaxy Evolution Explorer (GALEX) ultraviolet images.
We have conducted 
statistical analyses of the complete sample
of SB\&T galaxies, comparing with optical data and with control samples of normal galaxies
(Smith et al.\ 2007, 2010; Smith \& Struck 2010).  We have also conducted
detailed spatially-resolved
studies of several individual galaxies in the sample, comparing with numerical
simulations of the interactions
(Smith et al.\ 2005, 2008; Hancock et al.\ 2007, 2009; 
Peterson et al.\ 2009).
The current paper is a followup to an earlier study, 
in which we used Spitzer mid-infrared images (3.6 $\mu$m, 
4.5 $\mu$m,
5.8 $\mu$m, 8.0 $\mu$m, and 24 $\mu$m) and a ground-based
H$\alpha$ image to investigate star formation
in Arp 107 (Smith et al.\ 2005, hereafter \citetalias{smi05}).

In the \citet{arp66} Atlas optical photograph, 
Arp 107 is characterized by a strong spiral arm in the southern part 
of the larger galaxy, a tidal tail to the northwest, 
and a bridge towards a small elliptical companion in the northeast. 
The larger galaxy in the pair has a Seyfert 2 nucleus \citep{kee85}.  
In the Spitzer images, prominent knots of star formation are visible to the north
and northwest;
in combination with the spiral arm to the south this produces 
a ring-like appearance in the mid-infrared \citepalias{smi05}.
In \citetalias{smi05},
we presented a numerical simulation of the Arp 107 interaction, which 
demonstrated that 
Arp 107 is the product of an off-center collision, with parameters 
between those of a collisional ring and a prograde tidal encounter.
In our earlier study, we found a clear azimuthal sequence in the 
mid-infrared [3.6 $\mu$m] $-$ [8.0 $\mu$m] and [4.5 $\mu$m] $-$ [5.8 $\mu$m]
colors along the arm/ring.
Assuming that the 3.6 $\mu$m and 4.5 $\mu$m bands are rough tracers of the older
stellar population
and the 5.8 $\mu$m and 8.0 $\mu$m emission is dominated by 
interstellar particles
heated by young stars, these
color sequences 
suggest a sequence in the ages of the stars along the arm/ring, or,
alternatively, in the relative amounts of young and old stars.
In \citetalias{smi05},
we suggested that the pattern of stellar ages around the spiral arm may be due to 
differences in the time of maximum compression in the spiral arm.

In the current paper, we further investigate star formation along
the tidal arm of Arp 107,
by adding UV and optical images to this analysis.  
The hybrid morphology of this structure, between a collisional ring
and a tidal tail, 
strongly 
constrains its origin and evolution, and therefore 
it provides a good laboratory to study star formation triggering.
Throughout this paper, we assume a distance of
138 Mpc for Arp 107, using $H_{0}$ = 75 km~s$^{-1}$~Mpc$^{-1}$.  
In Section 2 of this paper, we describe the datasets, and in Section 3 we describe the
photometry.  In Section 4, we look at UV/optical/IR spectral energy
distribution (SED) plots for star-forming regions within
Arp 107.   In Section 5, we plot
the UV/optical/IR colors against position angle around the ring/arm.
In Section 6, we use single-burst
stellar population synthesis models to estimate
stellar ages and attenuations in these clumps.  We plot
these ages against azimuthal angle around the ring in Section 7.   
In Section 8, we explore two-component population synthesis models,
while in Section 9, we investigate the diffuse starlight in the ring.
A comparison is made to numerical simulations of the 
encounter in Section 10.  
In Section 11, we compare with results for other galaxies,
and summarize the results in Section 12.

\section{Data and Morphology}

In the current paper,
we utilize GALEX\footnote{http:$//$galexgi.gsfc.nasa.gov$/$docs$/$galex} 
far-ultraviolet (FUV) and near-ultraviolet
(NUV) images and Sloan Digitized Sky Survey 
(SDSS\footnote{http:$//$www.sdss.org$/$links.html}) 
ugriz optical images of Arp 107.
These images
were previously presented in a survey paper of the full SB\&T sample \citep{smi10}.
The GALEX images have a pixel size of 1\farcs5 and a circular field of view with a diameter of 1.2 degrees.  The NUV band covers 1750 $-$ 2800 \AA, while the FUV band covers 1350
$-$ 1705 \AA.  
The total GALEX exposure time in the NUV band was 2610 seconds, and 
in the FUV band it was 1094 seconds \citep{smi10}.
The spatial resolution of the GALEX images is approximately 5$''$.
The SDSS images have a pixel size of 0\farcs4, a field of view of 
13\farcm5 $\times$ 9\farcm8, and a point spread function (PSF) 
full width half maximum
(FWHM) of $\sim$1\farcs3.  The SDSS u, g, r, i, and z filters have effective wavelengths 3560, 4680, 6180, 7500, and 8870 \AA, respectively.
Further details on the GALEX and SDSS observations and
data are available in \citet{smi10}.

As described in \citetalias{smi05}, the 
Spitzer\footnote{http:$//$ssc.spitzer.caltech.edu} data 
consists of Infrared Array Camera (IRAC) images in the 3.6, 4.5, 5.8, and 
8.0 {$\mu$}m broadband filters, as well as 
a 
Multiband Imaging Photometer
(MIPS) image in the 24 $\mu$m broadband filter.   
The IRAC images
have pixel sizes of 1\farcs2 and a field of view of 
9\farcm0 $\times$ 14\farcm8, while
the MIPS
image has a field
of view of 7\farcm4 $\times$ 8\farcm1 and a pixel size of 2\farcs45.
The FWHM PSF is 1\farcs7 $-$ 2\farcs0 for the IRAC images and 6$''$
for the 24 $\mu$m image.
The H$\alpha$ image was obtained with the 0.6 m Erwin Fick Telescope in 
Boone, Iowa.  It has a field of view of 17\farcm9 $\times$ 17\farcm9, and
a pixel size of 1\farcs05.
Further details on these data are available in \citetalias{smi05}.

The SDSS g band optical image of Arp 107 is presented in Figure 1
(upper left panel),
along with the Spitzer 8 $\mu$m image (upper right panel).
An overlay of the Spitzer 8 $\mu$m image on the g image
is also provided in Figure 1 (lower left panel).
From the available images, it is unclear whether the tail to
the northwest is a continuation of the southern arm, or whether
it is part of another arm.
In the optical images, there is
a bend where the southern arm connects to the tail.
The bend may be due to the tail warping out of the plane of the disk;
alternatively, it may be due to the tail originating from elsewhere
in the system rather than the southern arm.  
In the 8 $\mu$m image, there is faint arm-like structure
extending westward out from the nucleus towards the tail.
The tail
may be the extension of this second arm rather than of the southern arm.
Velocity maps may provide useful information about this issue.

At the base of the strong arm in the east
there are several filamentary structures 
extending to the east visible in the SDSS images.
The bridge connecting the two galaxies appears double in the SDSS
images, as does the southern portion of the ring.
The `double' structure of the southern ring is also visible
in the lower resolution 3.6 $\mu$m Spitzer image.

\section{Photometry}

In \citetalias{smi05},
we identified 29 `clumps' of 
emission in and near Arp 107
using the Spitzer 8 $\mu$m image.
These clumps are labeled in the lower right panel of Figure 1.
Coordinates of these clumps are given in Paper I.
These are mainly knots
of star formation in the Arp 107 system, but also include the two
galactic nuclei, some likely background
objects, and a known foreground star.
Some of these clumps also show up as discrete knots in the
SDSS images (clumps 5, 7, 10, 17, 21, 26, and 27),
while some do not (clumps 4, 6, 8, 15, and 29).  
Clump 24 in the north lies about 5$''$ east of a bright optical
source.  
Clump 29,
near the end of the tail, does not have a discrete optical
counterpart, however, an optical knot is visible about 10$''$ 
further south
in the tail that does not have an 8 $\mu$m counterpart. 
Clump 21 lies at the
`bend' in the arm, where it intersects with the tail.
In the H$\alpha$ map presented in \citetalias{smi05},
only the brightest knots
in the ring were reliably detected 
(clumps 4, 5, 7, 10, 16, 21, and 26), with a calibration uncertainty
of $\sim$30\%.

In \citetalias{smi05}, we presented Spitzer and H$\alpha$ photometry of 
these 29 clumps.
We now provide GALEX UV and SDSS optical
measurements as well.
The 
aperture photometry was done with the IRAF\footnote{Image
Reduction and Analysis Facility; http:$//$iraf.noao.edu} {\it phot} command, 
using an aperture radius of 5$''$. We used a sky annulus with 
the mode sky fitting algorithm, 
an inner radius of 6$''$, and an outer radius
of 18$''$. 
Aperture corrections were done manually for the GALEX and SDSS images by taking counts of four moderately bright point sources
in the field and slowly increasing the radius until the counts leveled off.  For the FUV and NUV wavelengths the aperture 
corrections were multiplicative factors of 1.64 $\pm$ 0.19 and 
1.37 $\pm$ 0.10, respectively.
The SDSS filters were found to need no aperture corrections.
The SDSS photometry was corrected for Galactic reddening as 
in \citet{schlafly11}, as provided by the NASA
Extragalactic Database (NED\footnote{http://ned.ipac.caltech.edu}).
These corrections are 
very small (E(B $-$ V) = 0.025; E(g $-$ r) = 0.029).
The GALEX photometry was corrected
for Galactic reddening
using the \citet{cardelli89} attenuation law, which gives E(FUV $-$ NUV)
= 0.0044 and E(NUV $-$ g) = 0.096; alternative laws
\citep{fitzpatrick99, seibert05, yuan13} give values that differ
by $\le$0.1 magnitudes.
The final corrected photometry is given in Table 1.
Only statistical uncertainties from the IRAF {\it phot} routine
are included in Table 1; the quoted values 
do not include errors in the background determination, in calibration, and in 
the aperture
corrections.

As a test of this photometry, 
we obtained colors
for a 10\farcs4 $\times$ 25\farcs7 rectangular region
on the southern portion of the ring
containing clumps 4, 5, and 6.  
For sky subtraction, we used 
sky values obtained from multiple rectangular regions
far from the galaxies.
These three clumps
are relatively crowded, thus would be the most likely to
have problems with the photometry.
The colors for this rectangular region are consistent within
the uncertainties with those of the three clumps,
with the exception of the NUV $-$ g color
which is 0.5 magnitudes redder.  This is likely due to
the inclusion of some diffuse interclump emission in the
rectangular region, since the interclump light is 
considerably redder in this
color than the clumps (see Section 9).

\section{SED Plots}

For each of the 29 clumps, we created UV/optical/IR spectral energy distribution
(SED) plots,
in units of erg~s$^{-1}$~cm$^{-2}$ ($\nu$F$_{\nu}$).
These are
provided in the Appendix of this paper (Figures 13 $-$ 20).
Such plots help to categorize the clumps into two main groups, those with evidence of recent active star formation, or starbursts, and those with a more quiescent profile indicative of an older stellar population. 
A quiescent distribution has an approximately blackbody curve from the combined 
light of the stars, smoothly sloping up from lower UV values through
the optical, peaking at 1 $-$ 2 $\mu$m,
then smoothly sloping down towards the mid-infrared.
A starburst profile shows excess emission in the UV and mid-infrared
above the curve for an older stellar population.
The excess in the 5.8, 8.0, and 24 {$\mu$}m filters originates from interstellar grains and molecules heated mainly by young stars.  Excess emission in the UV is from recently formed hot stars. 

In this paper, we focus mainly on the clumps within the ring/arm.
However, for comparison, in the Appendix we provide SED plots for the other clumps
as well, some of which may not be associated with Arp 107.
As mentioned in \citetalias{smi05},
Clump 9, which lies inside of the 
ring, has an optical spectrum indicative of a foreground 
star, and its redshift shows it is not part of 
Arp 107 (W. Keel 2005, private communication).
As discussed in Paper I, Clumps 19 and 23, to the west of the system, 
have mid-infrared colors similar to that of quasars,
thus may be background objects.
The UV/optical/IR SED of Clump 23 (Figure 18) is roughly a power law, inconsistent with
that of a star formation region but supporting its identification
as a background quasar.   The SED of Clump 19 is inconclusive (Figure 17).

With the exception of clump 20, which is ambiguous,
the SED plots of all of the ring clumps
are consistent with on-going star formation.
In contrast,
the nucleus of the companion (Clump 28) has a SED profile indicating
an older stellar population.   The SED of the nucleus of the 
primary galaxy (Clump 14) also suggests an older stellar population,
but with some UV and mid-infrared excess, perhaps powered in part by the Seyfert nucleus.

Outside of the ring, clumps 24 and 29 have star-forming SEDs.
Clump 29 lies near the tip of the northwestern tail, thus is likely
part of Arp 107, while Clump 24
may be either a knot associated with Arp 107 or a background galaxy.
Clumps 12 and 25, 
also outside the ring, have 
relatively quiescent SEDs, suggesting that they might 
be faint foreground stars like Clump 9, 
however, clump 12 appears somewhat extended in the SDSS images, with
a FWHM of $\sim$1\farcs6, compared to $\sim$1\farcs3 for stars in
the field and Clump 9 and 25.
Clumps 2, 3, and 27, also outside of the ring, have relatively quiescent UV/optical SEDs,
but with possible 8 $\mu$m excesses.
Inside the ring, 
Clumps 11, 13, and 18
have star-forming SEDs.
Clumps 1 and 18 were detected in too few filters to classify.  

As seen in Figures 13 $-$ 20, for most of the clumps 
the SED 
drops between
8 $\mu$m and 24 $\mu$m, implying relatively
quiescent interstellar radiation
fields (ISRFs).  In general, the stronger the ultraviolet
ISRF, the higher the 24 $\mu$m emission
compared to that at 8 $\mu$m \citep{li01, peeters04, lebouteiller11}.
The 24 $\mu$m emission arises mostly from `very small interstellar
dust grains' (VSGs)
heated predominantly by UV light from O stars (e.g., \citealp{li01}).
The 8 $\mu$m Spitzer band contains emission from both 
very small dust grains
and polycyclic aromatic
hydrocarbons (PAHs).  Since PAHs
may be excited 
by non-ionizing photons
as well, the 8 $\mu$m emission
may be powered in part by lower mass
stars \citep{peeters04, calzetti07, lebouteiller11}.
The [8.0] $-$ [24] colors for the clumps in the Arp 107 ring
are less than 2.5 (Paper I), implying an ISRF
less than
or equal to about 10 times that in the solar neighborhood
\citep{li01}.  
As noted in Paper I, the 5 $-$ 8 $\mu$m surface brightness
in the Arp 107 disk clumps is a relatively low value of about 10 
L$_{\sun}$pc$^{-2}$, implying an ISRF approximately twice
that of the solar neighborhood. 
In contrast, for clump 14 (the Seyfert nucleus),
the SED increases strongly between 8 and 24 $\mu$m,
likely due to heating by the active nucleus.

The SEDs and their best-fit population synthesis models
are discussed further in Sections 6 and 8.

\section{Position Angle vs. Color Along the Arm}

In this section, we focus on the UV/optical/IR colors for the clumps
along the arm/ring.
In Figures 2 and 3, we plot various UV/optical/IR colors as a function
of position angle along the arm.
Note that these plots include clump 26 in the north near the bridge, which may not
be part of the primary arm.
As shown previously in Paper
I, there is a sequence in [3.6] $-$ [8.0] around the ring (see top panel
left in Figure 2).  The [3.6] $-$ [8.0] color is lowest (bluest)
in the east (PA = 20\deg $-$ 160\deg), increases towards the south
(PA = 180\deg $-$ 200\deg; clumps 4, 5, and 7), and is highest (reddest)
in the west
and northwest (PA = 240\deg $-$ 340\deg; clumps 10, 16, 21, and 26).
In \citetalias{smi05}, we interpreted this sequence as a sequence in the proportion
of young stars (as traced by the 8 $\mu$m band) relative to the older
stellar population (as measured by 3.6 $\mu$m).
A similar trend is seen in the [3.6] $-$ [24] color (top panel on right in Figure 2),
except that the clumps to the northeast (PA = 20\deg $-$ 80\deg; clumps 15, 17, and 22)
are redder than those in the south (PA = 120\deg $-$ 220\deg).

In Figure 2, we also provide the apparent 3.6 $\mu$m 
magnitude around the ring (bottom right panel). 
Assuming that the 3.6 $\mu$m emission is a rough estimate
of stellar mass, the clumps have relatively constant stellar mass, with
the exception of the fainter clumps 16, 20, and 22, and the slightly brighter
clumps 4, 5, and 6 to the south.
The apparent 24 $\mu$m magnitude is also plotted in Figure 2 (second panel from
the bottom on the right).   This gives an approximate measure of the star formation
rate in each clump.   
There is a very rough correlation of 24 $\mu$m luminosity (i.e., apparent
magnitude) with azimuthal angle, but with a lot of scatter.
Clumps 10, 21, and 26 in the west and north
have the highest star formation rates in the ring.

In FUV $-$ NUV and u $-$ g (Figure 2, bottom panel left and second panel right), 
the trend seen in the mid-infrared colors is reversed, with bluer optical colors
for the clumps with redder [3.6] $-$ [8.0] colors, and vice versa.
In NUV $-$ g and g $-$ r (Figure 2, second and third panels on left),
the southern clumps are again somewhat redder than the clumps in the west
and northwest, but there is more scatter.
As with the IR colors,
the UV/optical colors suggest either a sequence in age around the ring, or,
alternatively, a sequence
in the relative proportion of young and old stars.

The UV/optical colors, however, are affected by dust attenuation 
as well as age.
To get an estimate of how the dust attenuation 
varies around the ring, in Figure 3
we plot various tracers of dust attenuation 
against position angle around the ring.
In the top panel we have plotted
the H$\alpha$ to 8 $\mu$m
luminosity ratio, while the second panel
gives L$_{H{\alpha}}$/L$_{24{\mu}m}$ against position angle.
We have used the definition luminosity L = $\nu$F$_{\nu}$ for the Spitzer bands.
The lower two panels in Figure 3 show
FUV $-$ [8.0] and FUV $-$ [24] against position angle.
The luminosities in the FUV, H$\alpha$, 8 $\mu$m, and 24 $\mu$m bands are all
approximate tracers of young stars, with caveats.
The FUV and H$\alpha$ fluxes are also affected by dust attenuation.   
These tracers are also sensitive to the age of the stars.
The FUV is sensitive to both young and 
intermediate age stars ($\sim$100 Myrs), while H$\alpha$ 
traces younger
stars ($\le$10 Myrs).
As noted earlier, the 8 $\mu$m band may be powered
in part by older stars as well.

All four tracers of dust attenuation indicate that
the knots
in the northwest
(clumps 21 and 26, PA = 300\deg $-$ 360\deg) have higher dust attenuations
than the clumps in the west (clumps 10 and 16).
The clumps in the south (clumps, 4, 5, and 6 near PA 180\deg), are 
also more extincted than those in the west (clumps 10 and 16).
In comparing the southern clumps (clumps 4, 5, and 6) with the northwest clumps
(clumps 21 and 26), there are inconsistencies between the different
tracers of dust attenuation.   The southern clumps have lower
H$\alpha$ to mid-infrared
ratios 
than those in the northwest, 
indicating more dust attenuation,
but bluer FUV $-$ mid-infrared colors.  This suggests
some contributions from intermediate age stars to the observed UV light 
of the southern clumps, causing them to be blue in FUV $-$ mid-infrared
in spite of their relatively low
H$\alpha$ to mid-infrared ratios.
We return to this point in Sections 6 and 8.

\section{Population Synthesis:
Single-Burst Instantaneous Models}

We next compared the UV/optical colors of these clumps with 
population synthesis models
to determine the ages of the stellar populations within the clumps and their 
dust attenuations.
In this initial analysis we assume a single instantaneous burst;
in a later section (Section 8), we investigate more complex scenarios.
As in our earlier studies
\citep{smi08, hancock09},
we use Starburst99 version 
5.1 \citep{lei99} and include the Padova asymptotic giant
 branch stellar models \citep{vaz05}, assuming solar metallicity.
We assume 
a Kroupa initial mass function,
and integrate the model spectra
over the GALEX and SDSS bandpasses.
In the model spectra,
we 
include
the 
H$\alpha$ line from Starburst99 
as well as other optical emission lines, 
derived using the prescription given by \citet{anders03} for
solar metallicity star forming regions.
We used the \citet{calzetti94} 
starburst dust attenuation law.
As an example,
in Figure 4 we compare model FUV $-$ NUV vs.\ g $-$ r colors
with the observed colors of our clumps.

We computed ages, dust attenuations, and uncertainties on the ages and 
attenuations
using a chi squared ($\chi$$^2$) 
minimization calculation as in \citet{smi08} and \citet{hancock09}:
                                                                                     
$$\chi^2 = \sum^{N}_{i=1}\left(\frac{obs_{i} -
model_{i}}{\sigma_{i}}\right)^2$$

\noindent In this equation, N is the number of colors used in the analysis,
obs$_i$ is the observed color,  model$_i$ is the corresponding model
color, and $\sigma_i$ is the uncertainty in the obs$_i$ color.
For these fits, we used 
the FUV $-$ NUV, NUV $-$ g, u $-$ g, g $-$ r, r $-$ i, and i
$-$ z colors, when available.
Only filters with reliable detections were used to calculate 
the ages; upper limits ($\le$3$\sigma$ detections) were ignored.
In calculating the $\chi$$^2$ values, in addition to the statistical
errors,
we included additional uncertainties in the colors 
due to background subtraction and the aperture corrections.
As an estimate of the uncertainty in the colors of the
clumps due to background subtraction,
we compared the colors obtained with the above sky annuli with those
obtained using sky annuli with an inner radius of 12$''$ and an outer
radius of 18$''$. 
The median values of these additional
uncertainties are 0.03, 0.03, 0.05, 0.03,
0.01, and 0.02 magnitudes 
for FUV $-$ NUV, NUV $-$ g, u $-$ g, g $-$ r, r $-$ i,
and i $-$ z, respectively.
In addition, we included uncertainties in the GALEX aperture corrections
to the FUV $-$ NUV and NUV $-$ g colors.
To estimate the uncertainties in the best-fit parameters, we used
the $\Delta$$\chi^2$ method \citep{press92} to determine
68.3\% confidence levels for the parameters.
The best-fit parameters are provided in Table 2 along with the 
reduced chi-squared, $\chi^2$/(N$-$2), where N $-$ 2 is
the degrees of freedom.  Table 2 also lists the colors used in the 
fits.  We excluded the two galactic nuclei, the two likely quasars,
and the foreground star from Table 2, as well
as clumps with less than three colors available.

In the SED plots in the Appendix, 
we include the best-fit single-burst model results 
(solid black curves) for these 
clumps.
To give an indication of the uncertainty in the models,
we also plot models with the
best-fit dust attenuations but an age 1$\sigma$ less than
the best fit age (blue dashed curve), or 
1$\sigma$ more than
the best fit age (red dotted curve).
These three models are all scaled to the SDSS g band flux of
the clump.
These plots and the $\chi$$^2$/(N$-$2) values in Table 2
give an indication of how well the models fit the data.
In general, for a good fit $\chi$$^2$ should be approximately equal
to N $-$ 2.
As seen in Table 2, for most clumps $\chi$$^2$/(N$-$2) $\sim$ 1 to 3,
thus in most cases the model is a reasonably good fit to the data.
One factor that may contribute to increasing the $\chi$$^2$ values is 
variations in the dust attenuation law from location to location,
depending upon the geometry and the nature of
the dust (e.g., 
\citealp{witt00, boquien09, pancoast10}).
In addition, if the clumps are resolved in the GALEX images (FWHM $\sim$ 5$''$
$\sim$ 3 kpc) and there are color gradients in the clumps, 
the uncertainties in the aperture corrections may be under-estimated.
We also ignore stochastic effects in estimating
the ages, which may increase the $\chi$$^2$,
particularly for the lower mass 
clumps (e.g., \citealp{cervino02, popescu10}).  
Most importantly, for some clumps it is possible that 
more than one generation of stars contributes
to the observed light.   
This point is discussed
further in Section 8. 

Generally, it is assumed that the Spitzer 3.6 $\mu$m and 4.5 $\mu$m
bands are dominated by starlight with little attenuation, 
but this may not always be the case.
Some starbursting low metallicity dwarf galaxies
show strong excesses in these bands
above the stellar continuum inferred from optical data;
this is likely the result of 
highly extincted young stars, and/or
emission from hot dust grains, nebular emission lines,
and/or the nebular continuum
\citep{smith09}.
The young star forming region
in the northern tidal tail of Arp 285 shows an
apparent excess in these bands
above the inferred starlight \citep{smi08}, as does
the overlap region between the two disks of the Antennae galaxy
\citep{zhang10} 
and some of the tidal star forming
regions studied by \citet{boquien10}.
In contrast, 
for most of the Arp 107 clumps, the 
3.6 $\mu$m and 4.5 $\mu$m fluxes agree with the best-fit 
population synthesis models within the uncertainties 
of the models
(see Figures 13 $-$ 20).  
For most of the clumps there is no strong evidence
for excesses in these bands in Arp 107 above the stellar continuum
inferred from the optical/UV light,
though with the available data we cannot rule this out.
The Arp 107 clumps are older and less `starbursty' than
those in many other interacting galaxies (see Section 11), 
thus they are less likely
to have excesses in these bands. 

\section{Trends Along the Arm}

\subsection{Age vs.\ Position Angle Along the Arm}

For the clumps along the arm, in the left panel
of Figure 5 we plot the best-fit age from Table 2
vs.\ position angle.
A sequence in age is visible along the arm, with the 
younger clumps in the west/northwest 
(about 20 Myrs for clumps 10 and 16, at PA = 240\deg $-$ 280\deg) and
older clumps in the southeast/east
(about 80 Myrs for clumps 4, 6, and 8, at PA = 140\deg $-$ 180\deg).
This relative sequence of ages is similar to that inferred from
the [3.6] $-$ [8.0] color (Figure 2).

We obtained a second estimate of the ages of these clumps
from the H$\alpha$ equivalent width assuming
an instantaneous burst.   These also show a 
trend in the ages along the arm, but the ages are generally 
younger than those derived from the broadband colors, 
with ages
between 8 $-$ 13 Myrs ($\pm$$^5_3$ Myrs)
in the south (clumps 4, 5, and 6), and 
6 $\pm$ 1 Myrs in the
west (clumps 10, 16, and 21).
A similar difference 
between H$\alpha$-derived ages and ages from broadband photometry
was
found for some star-forming knots in Arp 284
\citep{peterson09}.
Such differences suggest that more than one generation
of stars are present in these clumps, either due
to multiple bursts, sustained star formation, or an underlying
older stellar population.  This issue is discussed further
in Section 8.  

\subsection{Attenuation vs.\ Position Angle Along the Arm}

In the right panel of Figure 5, we display the derived reddening
E(B$-$V) from SB99 against
position angle for the clumps along the arm (filled squares).
The inferred reddening
is highest for clumps in the southern and northern portion
of the ring, 
and lower in the eastern and western parts of the galaxy.
The pattern of reddening around the ring roughly matches that implied
by the dust attenuation tracers in Figure 3, but with
a lot of scatter.

We obtained a second estimate of the reddening
from the L$_{H\alpha}$/L$_{24{\mu}m}$ ratios of these clumps,
using the \citet{kennicutt09} relation between 
L$_{H\alpha}$/L$_{24{\mu}m}$ 
and 
dust attenuation, along 
with the \citet{calzetti94} starburst attenuation law.
The derived E(B $-$ V) values are higher in the south
(between 0.21 $-$ 0.28 for clumps 4, 5, and 6)
than in the west and north (0.10 $-$ 0.13 for clumps 10, 16, 21, and 26),
with only upper limits in the east ($\ge$0.29 for clumps 15 and 17).
These E(B $-$ V) values are plotted in the right panel of Figure 5 (open circles).
Note that the E(B $-$ V) values determined from
L$_{H\alpha}$/L$_{24{\mu}m}$ are lower limits
for position angles between 0\deg $-$ 160\deg.  These values 
are consistent with the values determined using SB99 within
the SB99 uncertainties, with the exception of clumps
5 and 6 in the south, where they are 1.5$\sigma$ $-$ 2$\sigma$ lower.
We note that the H$\alpha$ and probably
most of the 24 $\mu$m emission arises from
interstellar matter
associated with
a younger stellar population, 
which may be more attenuated than the starlight
that dominates the broadband UV and optical
light (e.g., \citealp{calzetti01}).

We obtain a third estimate of dust attenuation
by extrapolating from the 8 and 24 $\mu$m flux
densities to the
far-infrared flux as in \citet{calzetti05},
and then
using the FIR/FUV ratio along with the
FUV $-$ i color to estimate the
FUV dust attenuation as in \citet{cortese08},
assuming the \citet{cardelli89} A$_{FUV}$/E(B$-$V) ratio.
The 
\citet{cortese08} model FIR/FUV ratios
and FUV $-$ i colors 
were derived 
using the \citet{bruzual03} population synthesis models, 
including 
energy balance between the dust emission in
the infrared 
and dust absorption in the UV/optical.
As noted by \citet{cortese08} and references therein,
the FIR/FUV method for estimating
dust attenuation is almost independent of dust geometry
and dust attenuation law, however, it is dependent upon the
age of the stellar population.  This is partially corrected
for by \citet{cortese08} by including a UV/optical color
in the fitting as a constraint on the stellar age.
However, they note that their models
are not applicable for systems with obscured star formation
embedded within a older less obscured population.

The E(B $-$ V) values from the FIR/FUV method
are plotted in Figure 5 (right) as a dotted line.
The attenuations obtained with
the FIR/FUV method are generally less than 
those determined from the population synthesis models, but 
for most clumps they agree
within the uncertainties of the population synthesis, 
with the exception
of the clumps in the southern portion of the 
ring (clumps 4, 5, 6, and 7) (see Figure 5 right).
This difference
may be caused by contributions from a second less obscured 
intermediate age component contributing to the UV light,
which would lower the attenuation derived from the FIR/FUV 
method relative to that determined from fitting
the UV/optical SED.

We also experimented with single-burst models in which 
we fixed the attenuation to the values obtained by
the FIR/FUV method, and fitted for the best-fit age assuming
a single-burst population.
As expected,
the derived ages are older ($\sim$200 Myrs
for clumps 4, 5, 6 and 7), due to the lower assumed attenuations.
Also as expected,
the $\chi$$^2$ values are significantly worse, particularly
in the southern portion of the ring.
This supports the idea that two stellar populations are present.

\section{Population Synthesis: Models With Two Populations}

\subsection{Overview}

Next, we investigated population synthesis models using two 
populations of stars with different ages and dust attenuations. 
In these models, we 
varied the age of both components, the attenuation
of the younger component, and the burst strength (the ratio
of the mass 
of young stars relative to older stars).
We limited the younger population to ages less than 50 Myrs,
and the older population to between 50 Myrs and 10 Gyrs (i.e.,
intermediate to old stars).
This adds additional parameters to the fitting process, and 
therefore makes finding a unique solution
with well-constrained parameters difficult.  
Given the limited number of colors available for fitting and 
the uncertainties on these colors, it is possible to find
more than one distinctly different
decomposition of the SED that provides a reasonable fit to the data.
In the decompositions given below,
for simplicity
we assumed instantaneous bursts for both populations, or
an instantaneous burst superimposed on 
continuous star formation.
It is possible that the true star formation history
of these clumps 
may be better represented by one or more sustained bursts or 
an exponential decay with time, however, with the currently
available data
we cannot differentiate between these possibilities.
For the attenuation of the older stellar population, 
we investigated two different scenarios.  First, we 
explored models in which 
the attenuation to the older population is negligible.
Second, we explored models in which the dust reddening
of the older stellar population E(B $-$ V)(old) = 
0.44 $\times$ E(B $-$ V)(young).  
This latter scenario
is suggested by the fact that, for starburst
galaxies, the reddening to the stellar continuum has
been found to be 0.44 times that of the ionized gas
(e.g., \citealp{calzetti01}).
The ratio of the attenuation of the young stars to that
of the old depends upon the geometry of the clumps,
including
the locations of the young stars, the old stars, and the dust.

In the following sections, we provide three possible two-component
SED decompositions for two of the Arp 107 clumps.
These demonstrate the diversity of possible solutions,
and illustrates the uncertainty in SED modeling when additional
parameters are added.

\subsection{Young + Old Populations}

In Figure 6, we provide decompositions of the SEDs of clumps 6 and 15
assuming two instantaneous bursts of star formation, one young and one old.
In these plots,
the solid black curve is the older unextincted component, 
the blue dotted curve
is the younger more extincted component, 
while the red dashed curve is the sum
of the two components.
In these decompositions,
the younger population dominates the observed light
in the UV, while the older population accounts for the majority
of the optical light.
For
clumps 6, the young population
is 8 Myrs old, the older population 1500 Myrs,
the reddening of the younger population is E(B$-$V) = 0.08, and
the reddening of the older population is assumed to be zero.
For
clumps 15, the young population
is 15 Myrs old, the older population 350 Myrs,
the reddening of the younger population is E(B$-$V) = 0.04, and
the reddening of the older population is assumed to be zero.
The burst strengths are relatively weak,
with the young/old stellar mass ratios being 0.0061 and 0.079 for
clumps 6 and 15, respectively.
These two-component decompositions
provide better matches to the shape of the UV/optical SEDs than
the single-component models plotted in Figures 14 and 16.   Note that
for clump 6 the older stellar component can account
for the 3.6 and 4.5 $\mu$m 
fluxes, while for clump 15 it cannot, suggesting either interstellar
contributions in these bands, 
a more complex star formation history, and/or
a range in attenuations.

If the reddening of the older stellar population
is instead required to be 0.44 $\times$ E(B $-$ V)(young)), then
the age of the older population decreases 
slightly  
to 1000 Myrs while the young/old mass
ratio decreases only very slightly
(0.0058).  The age and reddening of the young population remained
the same.
When E(B $-$ V)(old) is required to be
0.44 $\times$ E(B $-$ V)(young) 
for clump 15, 
the parameters are unchanged. 

The attenuations in these models
are consistent
with the values determined 
from the FIR/FUV ratios 
(see Figure 5).  However, the derived E(B $-$ V) values are 
lower than those obtained from the L$_{H\alpha}$/L$_{24{\mu}m}$ ratios
(E(B$-V$) $\ge$ 0.30 for clump 6 and 
E(B$-V$) $\ge$ 0.29 for clump 15).  

\subsection{Young + Intermediate Age Stars}

Alternative decompositions of the SEDs of clumps 6 and 15
are presented in Figure 7.
In these models, a highly obscured young stellar population is
added to an unobscured intermediate age population.  The obscured
young
population contributes significantly to the longer wavelength
light, accounting for most of the short wavelength Spitzer light,
while the intermediate age stars contribute significantly
at shorter wavelengths.
Such a situation might arise when successive generations
of stars have formed in close proximity, and stellar winds/supernovae 
have partially cleared out the dust from around the 
intermediate aged stars and/or these stars
have 
escaped from their natal molecular clouds, 
while the young stars are still 
deeply
embedded (e.g., \citealp{charlot00, calzetti01, conroy10}).

In the decompositions in Figure 7,
the young populations
are 6 and 8 Myrs old for clumps 6 and 15, respectively, 
the intermediate-age populations are 200 Myrs and 75 Myrs old,
and the reddening of the younger population is E(B$-$V) = 0.8 and 1.0.
In the decomposition
of clump 6 in Figure 7, most of the light in the optical and short wavelength
Spitzer bands comes from the 
obscured younger population, while the UV light is dominated by the intermediate
age
less obscured stars.
In the decomposition shown in Figure 7 for clump 15,
the majority of the observed light
in the UV and shorter wavelength optical (FUV $-$ r)
comes from an unobscured intermediate-age stellar population.
The obscured younger population contributes in the longer wavelength
optical, and  
dominates the emission in the 3.6 $\mu$m and 4.5 $\mu$m Spitzer 
bands.  
For both clump 6 and clump 15, the decompositions in Figure 7
can account for the observed 3.6 $\mu$m and 4.5 $\mu$m emission,
without needing to include interstellar emission in those bands.

Note that the models presented in Figure 7 are very different 
from those in Figure 6.  In Figure 6, the UV mainly comes from
young stars and the optical from older stars.
In contrast, in Figure 7, the UV is from intermediate age stars
and the optical from very obscured younger stars (clump 6)
or from young plus intermediate age stars (clump 15).
The young/old stellar mass ratios in these second sets of models
are much higher than those shown in Figure 6, 
being 0.79 for
clump 6 and 2.4 for clump 15.
As can be seen in Figure 7, for both
clumps the alternative decompositions
also give good matches to the photometric data.

Note, however, that the decompositions in Figure 7 are inconsistent with
the assumption that the 
reddening to the older stars 
is half that to younger stars; there is a much
larger difference in the attentuations to the two populations.
Further, the attenuations are larger than those derived from
either the FUV/FIR method or
the L$_{H\alpha}$/L$_{24{\mu}m}$ ratios.

We also experimented with decompositions in which 
we required 
that E(B $-$ V)(old) =
0.44 $\times$ E(B $-$ V)(young), 
searching for scenarios in which
the UV is dominated by light from 
an intermediate aged population and the longer-wavelength
optical light from a younger population.
With such assumptions,
a reasonable match to the SED of clump 15 can be made
with E(B $-$ V)(young) = 0.27,
E(B $-$ V)(intermediate) = 0.12, age(young) = 8 Myrs, 
and age(intermediate) = 50 Myrs,
with a young/intermediate mass ratio of 2.0.
This is consistent with the reddening to the younger population
inferred from the 
L$_{H\alpha}$/L$_{24{\mu}m}$ ratio.
It is higher than
that derived from the FIR/FUV ratio,
however, if the FUV arises from intermediate-aged
stars, and the  
mid-infrared fluxes (from which we infer the FIR flux) 
arises from dust associated with young stars, then the FIR/FUV
ratio is not necessarily
a good indicator of the attenuation to the young
stellar population.  

For clump 6
we were not able to get a good match
to the SED with the standard differential attenuation 
assumption and an intermediate
plus young stellar population (intermediate
age $\le$ 500 Myrs and E(B $-$ V) $<$
1.1).
Instead, the models tend to converge
to a single in-between age for both components.
This argues against the decompositions shown in Figure 7, if the 
assumption of E(B $-$ V)(intermediate) = 0.44 $\times$ E(B $-$ V)(young)
is correct.

\subsection{Continuous Star Formation + An Instantaneous Burst}

The observed SEDs are not well-matched by continuous
star formation models.  
However, we were able to get good matches to the SEDs by combining
continuous star formation with an instantaneous burst and
fixing the attenuation ratio to the \citet{calzetti01} value.
Examples are shown in Figure 8.
The model for clump 6 combines continuous star formation starting 10 Gyrs
ago with an instantaneous burst 100 Myrs old.  
The reddening to the
intermediate-age starburst is E(B $-$ V) = 0.22, and 
E(B $-$ V) = 0.10 to the other stars.
For clump 15, the decomposition in Figure 8 consists of continuous
star formation starting 10 Gyrs ago with an instantaneous burst 50 Myrs old.
The reddening to the
intermediate-age starburst is E(B $-$ V) = 0.10, and 
E(B $-$ V) = 0.04 to the other stars.
In these scenarios, the age of the burst
is not well-constrained, and it does not dominate the UV light.

The different SED decompositions shown in Figures 6, 7, and 8 illustrate
the range of possible models for these clumps.  Likely the true
decomposition is in-between these examples, with a range of stellar ages
and attenuations, perhaps with exponentially declining starbursts or
extended starbursts, rather than instantaneous or continuous star formation.  
It is also possible that three distinct populations may
exist in these clumps: a young burst, an intermediate-age burst,
and an older underlying stellar population.  In this case, the third
population may be hard to hard to discern observationally.
In any case, the SEDs suggest that two or more stellar populations
are present, though they are not well constrained.

To better determine the star formation histories of these clumps,
UV, optical, and/or near-IR spectra would be valuable,
to search for direct evidence for an intermediate age
stellar population, and to obtain alternative measures of the dust
attenuation
to the younger stars.
Near-infrared photometry in the 1 $-$ 3 $\mu$m range
would also be useful, since it
which would provide additional
constraints on the models.   
For example, in Figure 6b the model SED of clump 15
drops with increasing wavelength between the SDSS bands 
and 2 $\mu$m, while in Figure 7b the model SED rises, 
with both decompositions
providing a good match to the SDSS fluxes.
Higher spatial resolution observations would be also useful 
in determining the spatial
distribution of the young stars compared to the older stars.

\section{The Interclump Starlight}

Another way to determine the properties of the older stellar population
is to obtain the SED of the light arising 
from between the 8 $\mu$m clumps.   
To this end, we extracted fluxes in an annulus aligned along the ring,
as shown in 
Figure 9 (left panel).  This annulus
has an inner radius of 18$''$ and an outer radius of 34$''$, and excludes
the Seyfert nucleus and the foreground star (clump 9), as well
the clumps in the north (21, 22, 24, and 26) and those outside the ring.
For sky subtraction we used the mode sky-fitting algorithm and
a sky annulus with an inner radius
of 34$''$ and an outer radius of 44$''$.
The SED of the extracted light, minus the light from the 
clumps located in the annulus, is provided in Figure 9 (right panel).
This SED is relatively quiescent, with some excess in the UV and 
mid-infrared.
It is much more quiescent than that of the clumps in the ring, indicating that these
knots of star formation are embedded in a more diffuse older stellar population.
The NUV $-$ g color for the interclump light
(NUV $-$ g = 3.1) is considerably redder
than that of the clumps (NUV $-$ g from 0.5 to 1.8), 
thus the interclump stars are generally
older than the stars in the clumps.
The UV/optical colors of this interclump light is
in the middle of the range found for normal spiral galaxies
\citep{smith10}, except that the FUV $-$ NUV and i $-$ z colors are somewhat bluer
than average.   
The SED of this interclump light
resembles that of the two galactic nuclei (clumps 14 and 28), and, like those
regions, does not
fit a single-population instantaneous burst model well, but 
instead indicates
more than one age stars.   

An example two-component SED decomposition is over-plotted in Figure 9b,  
which matches the shape of the UV/optical SED well.  
This decomposition includes an 8 Myrs old burst reddened by E(B-V) = 0.08,
combined with an older population of age 1500 Myrs.  The young/old
stellar mass ratio in this decomposition is a relatively
low value of 0.0011.
Thus the arm/ring appears to have an older stellar
population which is more extended than the young stars.  
The star formation is clumpier than the older
stars, but is not completely
confined to the circular apertures used in this study.

\section{Comparison with Numerical Simulations of the Interaction}

In Paper I, we presented a numerical simulation of the Arp 107 
encounter, which reproduced the observed structure with
an off-center collision between a low mass early-type galaxy
and a disk.   The assumed mass ratio was set to 0.16, and
the impact occurred near the outer edge of the primary disk,
with the orbit of the companion being inclined to the rotational
plane of the primary.
In the early stages of this encounter, the primary had the
characteristic two-armed structure of a prograde encounter,
with one arm evolving into the observed bridge.  
This simulation is in-between that of a classical ring
galaxy, a head-on collision with
an orbit perpendicular to the primary's disk, 
and a prograde planar encounter with a large
impact parameter.
The impact parameter of the Arp 107 model is larger than that
used in simulations of the `Sacred Mushroom' system 
\citep{wallin94}, which shows a closed ring-like structure
rather than a tidal arm/tail.  Their Mushroom model had an impact
parameter of about half of the disk radius.
\citet{toomre78} presented a series of simulations of
ring galaxies with increasing impact parameters;
closed ring-like structures are produced in systems
with smaller impact parameters, while tail-like structures
are produced with larger values.

In \citetalias{smi05}, we suggested 
that the sequence in Spitzer colors along the arm was caused 
by a progression
in the time of maximum compression along this feature. 
One way to test this idea, and to test whether the ages 
derived from the 
single-burst model
are reasonable, is to compare the apparent rate of migration of the star formation
along the tidal arm with theoretical expectations.
We can estimate this rate
by 
comparing the ages of clumps at the base/middle of the tail 
(southeast/east) to those near the end of the 
tail (west).
These areas 
are separated by about 35 kpc.
If we use the single-burst ages, and
associate an age of around 80 Myrs to the base/middle 
(clumps 4, 6, and 8), and 20 Myrs at the end (clumps 10 and 16), 
we get a star formation migration rate of 0.58 kpc/Myr (567 km/s).

In the simulation presented 
in \citetalias{smi05}, which uses a local density-dependent
star formation algorithm, star formation does not 
travel smoothly along the arm, but occurs stochastically
throughout the arm.
However, there is a rough trend of the main 
locations of star formation
moving outward along the arm.
If we isolate a few frames in the model where star formation seems to
be most active in the base, middle, and tail of the arc, and calculate
a model star formation migration rate, it is much lower (more than
a factor of 10) than the rates calculated above.  

A comparison can also be made to the models 
of prograde planar encounters
presented by \citet{wal90}, who investigates density 
compression along tidal features.  In his Figure 3, he 
provides density vs.\ time for various locations along a 
tidal feature in a prograde planar interaction.  The density 
enhancement moves from the base of the tail to near the end 
in about 0.85 model time units, a distance of about 1 model unit.  
In his scale, this converts to roughly 0.21 kpc/Myr (205 km/s), 
about 3 times smaller than the rate we infer for the Arp 107 
tail from the instantaneous burst
models.

These inconsistencies suggest that the situation is more
complex than the simple picture of a compression wave traveling
along a tidal arm.    This is supported by our 
two-component decompositions of the SEDs, which suggest
that the clumps 
may have both young and old stars.
The observed azimuthal UV/optical/IR color sequences 
may be due to 
increasing proportions of young to old stars along the arm, 
rather than a sequence in the true ages of the stars along the arm.
Star formation may have occurred at about the same time at
all locations along the arm, likely triggered by the encounter.
In this sense, the system may have more in common with 
classical ring galaxies than with classical prograde planar encounters.
Unlike the rings in many classical ring galaxies, which appear to 
be mainly young stars \citep{higdon95, marston95, boquien07},
the Arp 107 arm/ring
has a substantial underlying intermediate
age
or older population.

\section{Comparison to Other Galaxies}

We next compared the properties of the Arp 107 clumps 
with those in other nearby pre-merger 
interacting galaxies from the literature.   In Figures 10 and 11, 
we provide two sets of UV/optical/IR color-color plots.  
In these plots, for clarity we only included the clumps in Arp 107 that
lie along the
strong arm/ring, and the clump at the end of
the tidal tail (clump 29).  
These plots also include clump photometry from Arp 82 \citep{hancock09},
Arp 285 \citep{smi08}, Arp 284 \citep{peterson09}, Arp 24 \citep{cao07}, 
Arp 244 \citep{zhang10}, and Arp 143 \citep{beirao09},
along with clump i from NGC 2207 \citep{elmegreen06}.
In Figure 11, we also included photometry from \citet{boquien10} for
extra-disk star forming regions in the interacting galaxies Arp 105, 
Arp 245, NGC 5291, NGC 7252, and VCC 2062.
Some of the more unusual clumps are identified in the figure captions.
We excluded galactic nuclei from these plots, along with the likely background
quasar near Arp 82 (see \citealp{hancock07}) and the Arp 143 clump
for which the light was affected by a foreground star \citep{beirao09}.
We also included photometry from
\citet{boquien09} for H~II regions in
five nearby normal spirals and the interacting system M51 (Arp 85).  
Since published photometry is not available for all of these systems in all of the bands
plotted in Figures 10 and 11, the plots are incomplete, with some sources
not appearing in all of the panels in these Figures.

There is considerable scatter
in the UV/optical colors plotted in Figures 10 and 11.
This scatter is likely caused by differences in the ages of the stars, varying
amounts of dust attenuation, as well as differences in the geometry of the star forming
region and in the dust properties.   
The FUV $-$ [24] colors correlate 
to some extent with [3.6] $-$ [24] (Figure 11, middle panel bottom row), with
younger clumps having redder FUV $-$ [24] colors, implying more dust absorption.
The Spitzer [3.6] $-$ [24] colors also correlate with [8.0] $-$ [24] (Figure 11,
lower right panel).
This indicates that the larger the proportion of young stars, the more emission
from hot VSGs compared to PAH emission.   
Arp 107 clumps 4, 5, 6, 7, and 8, along with the tail-end clump 29,
stand out in this plot and in the FUV $-$
NUV vs.\ [8.0] $-$ [24] plot (Figure 11, upper right) as having low [8.0] $-$ [24] colors.
This suggests that PAH emission in these clumps is important relative
to the VSGs, and may be powered in part by older stars.
The single-burst ages derived for these clumps of 55 $-$ 100 Myrs (Table 2)
are consistent with a large
population of B stars
but few O stars.
B stars may be important in PAH heating (e.g., \citealp{peeters04}), while O stars
may dominate VSG heating.

In Figures 10 and 11, Stephan's Quintet clump 5,
the Arp 82 hinge clump, the Arp 82 end-of-tail clump, 
and clump i in NGC 2207
stand out as being
very `starbursty', that is, having large [3.6] $-$ [24] ratios,
in contrast to the Arp 107 clumps.
This is consistent with the population synthesis of \citet{boquien10},
which indicates a very young population in SQ 5 (2 Myrs) with few underlying older stars.
The very large [8.0] $-$ [24] colors for these sources indicate 
very enhanced UV fields.
According to the \citet{li01} dust models, such large [8.0] $-$ [24] ratios are reached with
ISRFs about 10,000 times that of the local neighborhood.
Even the most active star forming regions in Arp 107 (clumps 10, 16, 21, and 26)
are much more quiescent than these starburst clumps. 
Interestingly, in Figures 10 and 11, the quiescent Arp 107 clumps 
lie near Stephan's Quintet clump 2 (open magenta circle), 
which has a similar derived age of 85 Myrs 
according to the population synthesis modeling of 
\citet{boquien10}.

Two other sources that stand out in these plots are 
VCC 2062 (cyan open hexagon) and 
clump 26 in the northern tail of Arp 82 (green asterisk).   They are somewhat quiescent
in [3.6] $-$ [24], but are quite red in [8.0] $-$ [24],
not fitting the general correlation between [3.6] $-$ [24] and [8.0] $-$ [24].   
They are also extremely blue in [3.6] $-$ [8.0] compared to the other clumps.
This suggests that these objects are deficient in PAH emission
relative to the other clumps in these plots.
In 
Spitzer [3.6] $-$ [4.5] vs.\ [4.5] $-$ [8.0] and [4.5] $-$ [5.8]  
vs.\ [5.8] $-$ [8.0] color-color plots,
clump 26 in the Arp 82 tail lies very close to the location expected for stars,
i.e., colors approximately equal to zero
\citep{hancock07}.  It was undetected in the \citet{hancock07} H$\alpha$ map,
and no redshift is available, thus whether or not it is part of the Arp 82 system
is uncertain.  
VCC 2062 is a dwarf galaxy in the Virgo cluster, lying within an HI structure that
extends to a nearby spiral, suggesting it may be a tidal dwarf \citep{duc07}.
From SED fitting, \citet{boquien10} estimate a burst
age of about 115 Myrs for VCC 2062, similar to or somewhat
older than the Arp 107 clumps.  

In Figure 12, we provide
another comparison of the Arp 107 clumps with the clumps in these other
galaxies.
In Figure 12, we plot the 24 $\mu$m luminosity ($\nu$L$_{\nu}$)
against the [3.6] $-$ [24] color.
There is a very weak correlation of [3.6] $-$ [24] color
with 24 $\mu$m luminosity, with a lot of scatter.  The scatter
may be due 
in part to the fact the physical sizes of the areas defined
as `clumps' 
varies from galaxy to galaxy in this sample.
As noted in Section 5, there is a rough trend of increasing 24 $\mu$m
luminosity and increasing [3.6] $-$ [24] color with increasing
azimuthal angle in Arp 107 (see Figure 2, second panel).
Figure 12 shows that the Arp 107 clumps are relatively quiescent
in [3.6] $-$ [24], and are not in the L$_{24}$ range of the hot spot in Arp 244,
clump i in NGC 2207, or the hinge clump in Arp 82.   
The Arp 244 hot spot and clump i in NGC 2207 stand out in this plot
as being both very
red in [3.6] $-$ [24], and as having very high 
24 $\mu$m luminosities. 
The most 24 $\mu$m-luminous knots
in Arp 107 are clumps 10 and 26, a factor of ten times lower
in L$_{24{\mu}m}$
than the Arp 244 hot spot and clump i in NGC 2207, and a factor of 
two lower 
than the Arp 82 hinge clump.  
The end of the Arp 107 tail is moderately luminous at both
24 $\mu$m and 3.6 $\mu$m, so 
is low in [3.6] $-$ [24].
The candidate tidal dwarf galaxy VCC 2062 and 
Arp 82 clump 26 stand out in this plot
as having
particularly low
L$_{24{\mu}m}$. 

These plots show that the Arp 107 clumps are
relatively quiescent compared to these other sources, 
being bluer in the mid-infrared
colors.  
In fact, the Arp 107 clumps extend the trends in several 
of the color-color plots 
considerably to the blue. 
This implies that either the stars are older,
or there are greater contributions
from an old population compared to young stars.
If this 
is indeed the result of greater 
stellar ages, then the 
Arp 107 ring/arm may contain some of 
the oldest clusters discovered 
so far in a tidal structure. 

\section{Summary}

In Paper I, strong azimuthal variations in the mid-infrared
colors of clumps were seen in the ring/primary arm of Arp 107.
In the current paper, we find corresponding trends in the 
UV/optical colors along this structure, with bluer colors
further along the arm.
Single-burst population synthesis modeling reveals 
a trend in average stellar age of the star-forming clumps
along this arm, from
around 80 Myrs old near the base of the arm, decreasing to
around 20 Myrs near the end. 
However, two-component population synthesis models
suggest that the situation is more complex.  Likely
all of the clumps have some young stars with ages
around 10 Myrs, along with some
intermediate age and/or older stars.
Thus the observed sequences in the UV/optical/IR colors
along this feature are likely due to variation in the proportions
of young to old stars along this arm, rather than
in the stellar ages themselves.
The clumps in Arp 107 are somewhat quiescent compared to those
in many other interacting galaxies.

Given the uncertainties in the data, it is difficult to 
distinguish between alternative two-component decompositions
of the SEDs for the clumps along this arm.
To better determine the stellar populations in this system, 
followup near-infrared photometry
and UV/optical/IR spectroscopy would be valuable.
Higher spatial resolution observations would be useful in determining the
spatial distribution of the young stars compared to the older stars.

This research was supported by NASA Astrophysics Data Analysis Grant ADAP10-0005 and NASA Spitzer Grants RSA 1353814 and 1379558.  This research has made use of the NASA/IPAC Extragalatic Database (NED), which is operated by the Jet Propulsion Laboratory, California Institute of Technology, under contract with NASA.
We thank Mark Hancock for the use of his scripts to determine
population ages, and Mark Giroux for helpful 
comments.
We thank the anonymous referee for helpful suggestions.

\section{Appendix}

In this Appendix we present SED plots for all 29 clumps.  These are displayed in 
Figures 13 $-$ 20.

\begin{figure}
\plotone{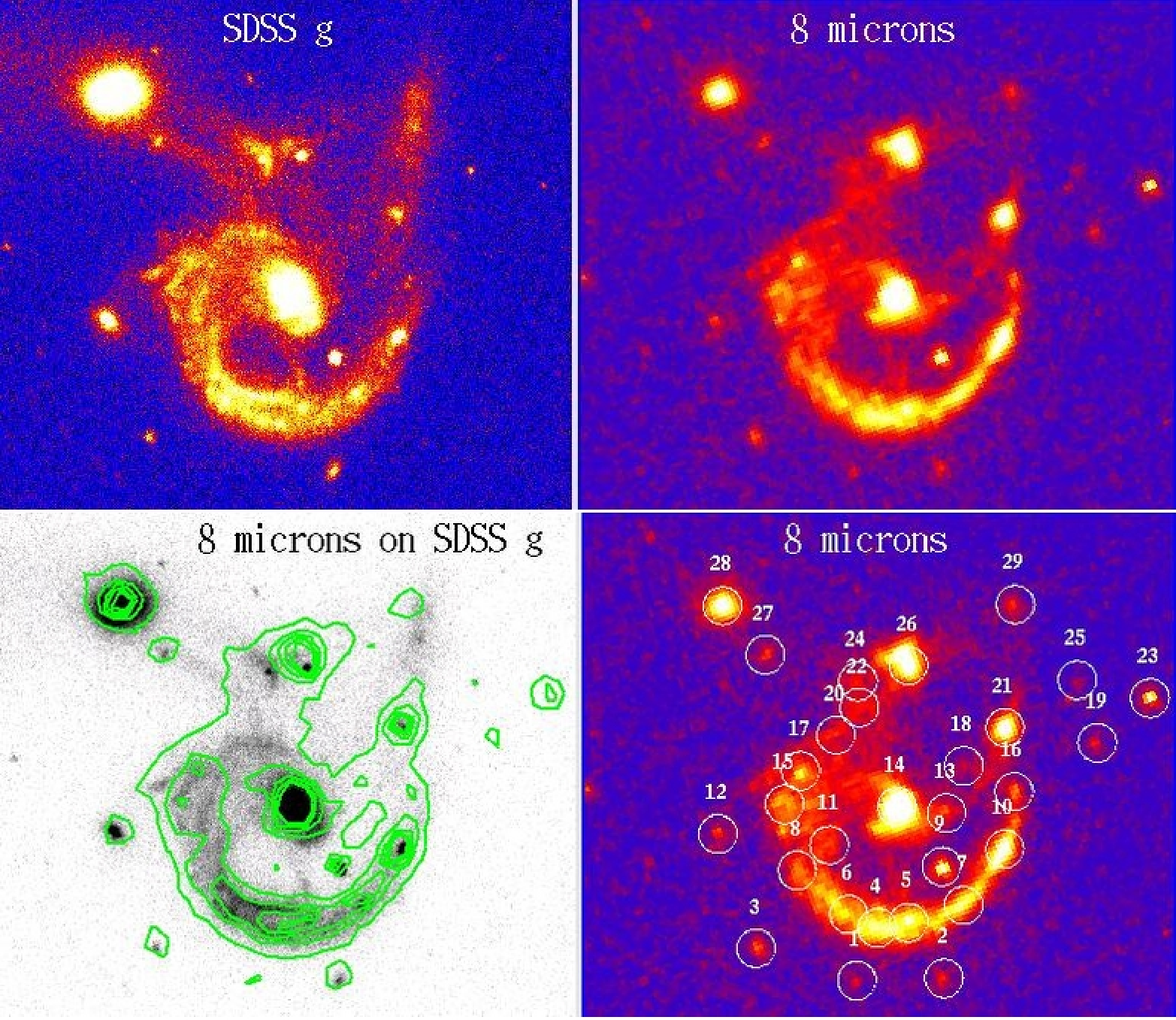}
\caption{Top left: The SDSS g band image of Arp 107.
Top right: 
The Spitzer 8.0 {$\mu$m} image of Arp 107.
Bottom left: The 8.0 $\mu$m contours superimposed on the SDSS g image.
Bottom right: The 8.0 {$\mu$m} image of Arp 107, with the 29 clumps marked
with 5$''$ radius circles.  
The clumps are numbered in order of increasing declination. 
North is up and east to the left in these pictures.
The field of view is 2\farcm3 $\times$ 2\farcm3.  The nucleus of the main
galaxy is at 10$^{\rm h}$ 52$^{\rm m}$ 15$^{\rm s}$, +30$\deg$ 3$'$ 28$''$
(J2000).
Coordinates for the other clumps are given in Paper I.
}
\label{fig1}
\end{figure}

\begin{figure}
\plottwo{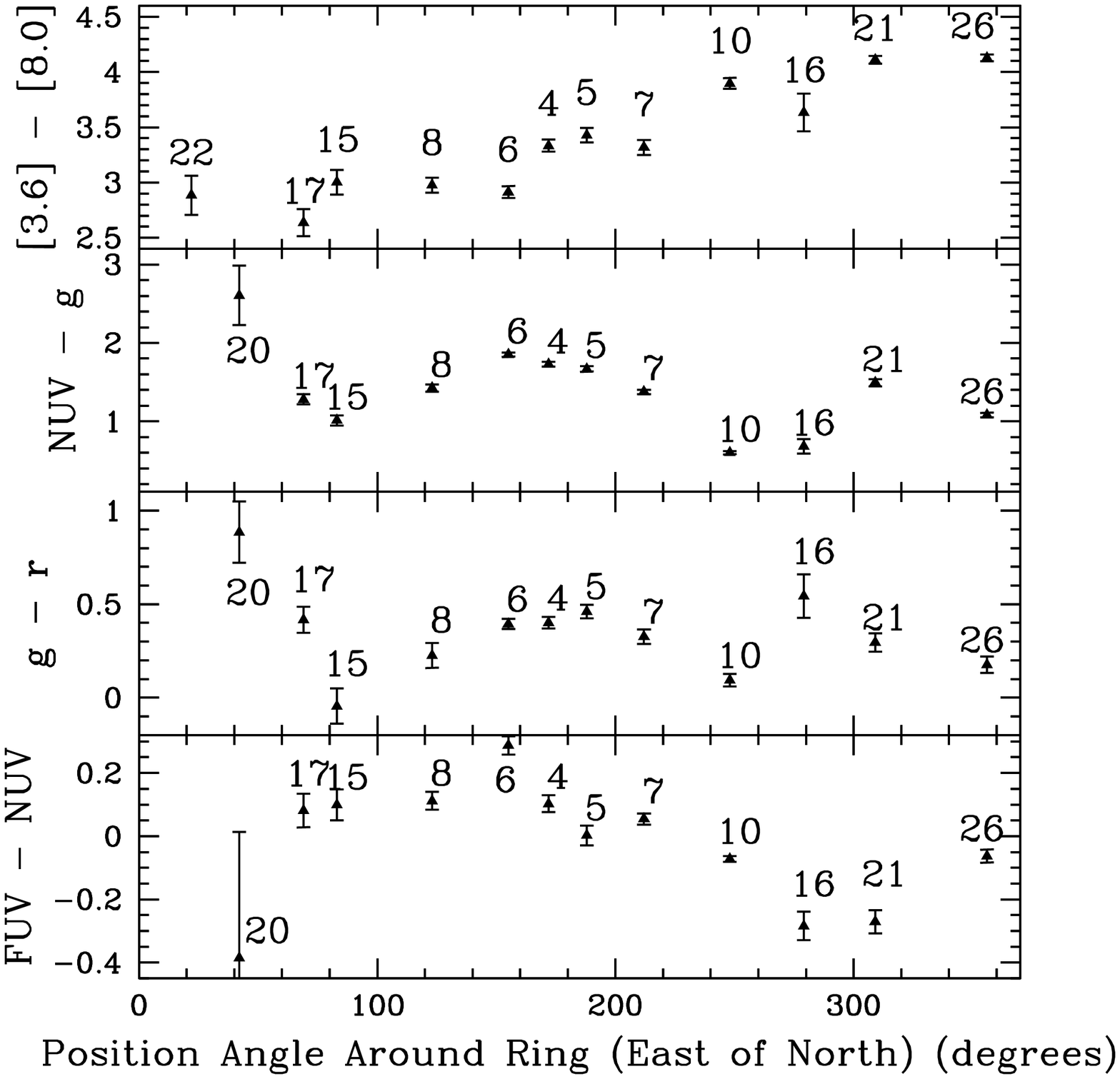}{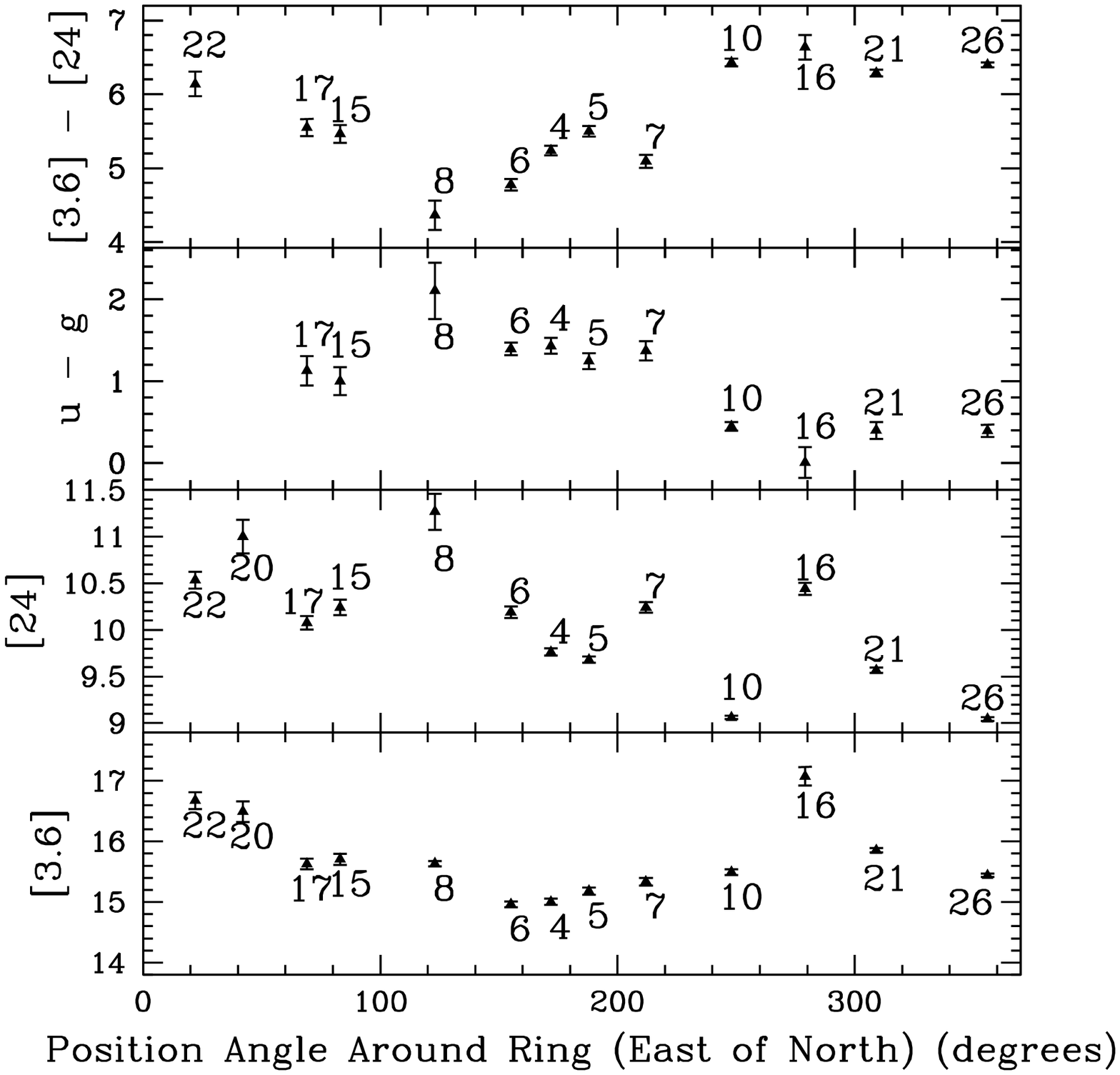}
\caption{Various UV/optical/IR colors for clumps in the spiral arm as a function of 
position angle, east of north. The locations of these clumps are marked in Figure 1. 
}
\label{fig2}
\end{figure}

\begin{figure}
\plotone{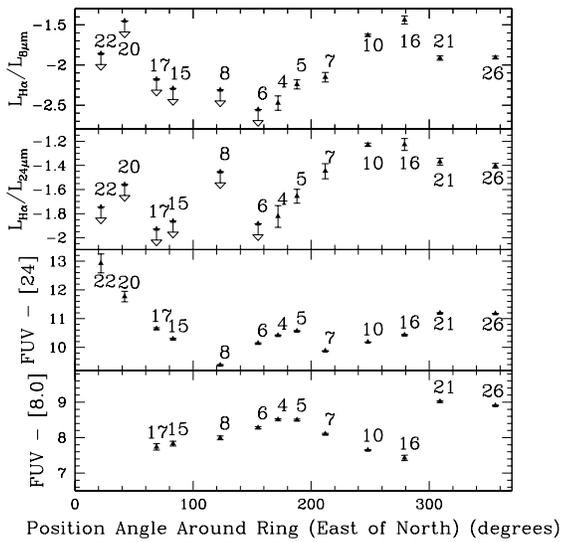}
\caption{Additional UV/optical/IR colors for clumps in the spiral arm as a function of 
position angle, east of north. 
}
\label{fig3}
\end{figure}

\begin{figure}
\includegraphics{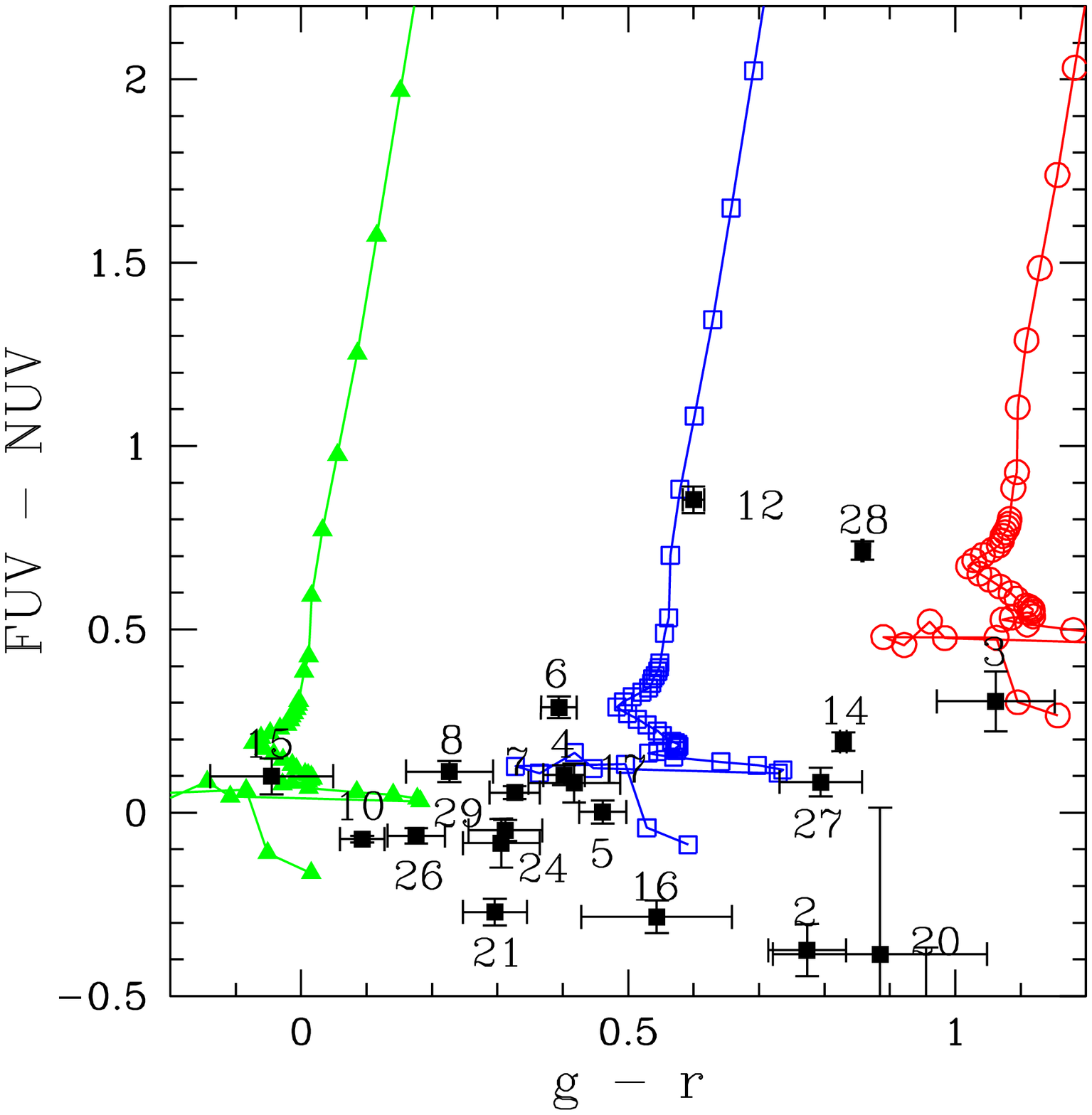}
\caption{The green triangles show an instantaneous population synthesis model with E(B-V) = 0, the blue squares show E(B-V) = 0.5, and the red squares show E(B-V) = 1.0. The model ages start at an age of 1 Myr, then increase by 1 Myr steps to 20 Myr, then by 5 Myr steps to 50 Myr, then 10 Myr steps to 100 Myr, 100 Myr steps to 1 Gyr, and 500 Myr steps to 10 Gyr. The Arp 107 clumps are labeled as in Figure 1.}\label{fig4}
\end{figure}

\begin{figure}
\plottwo{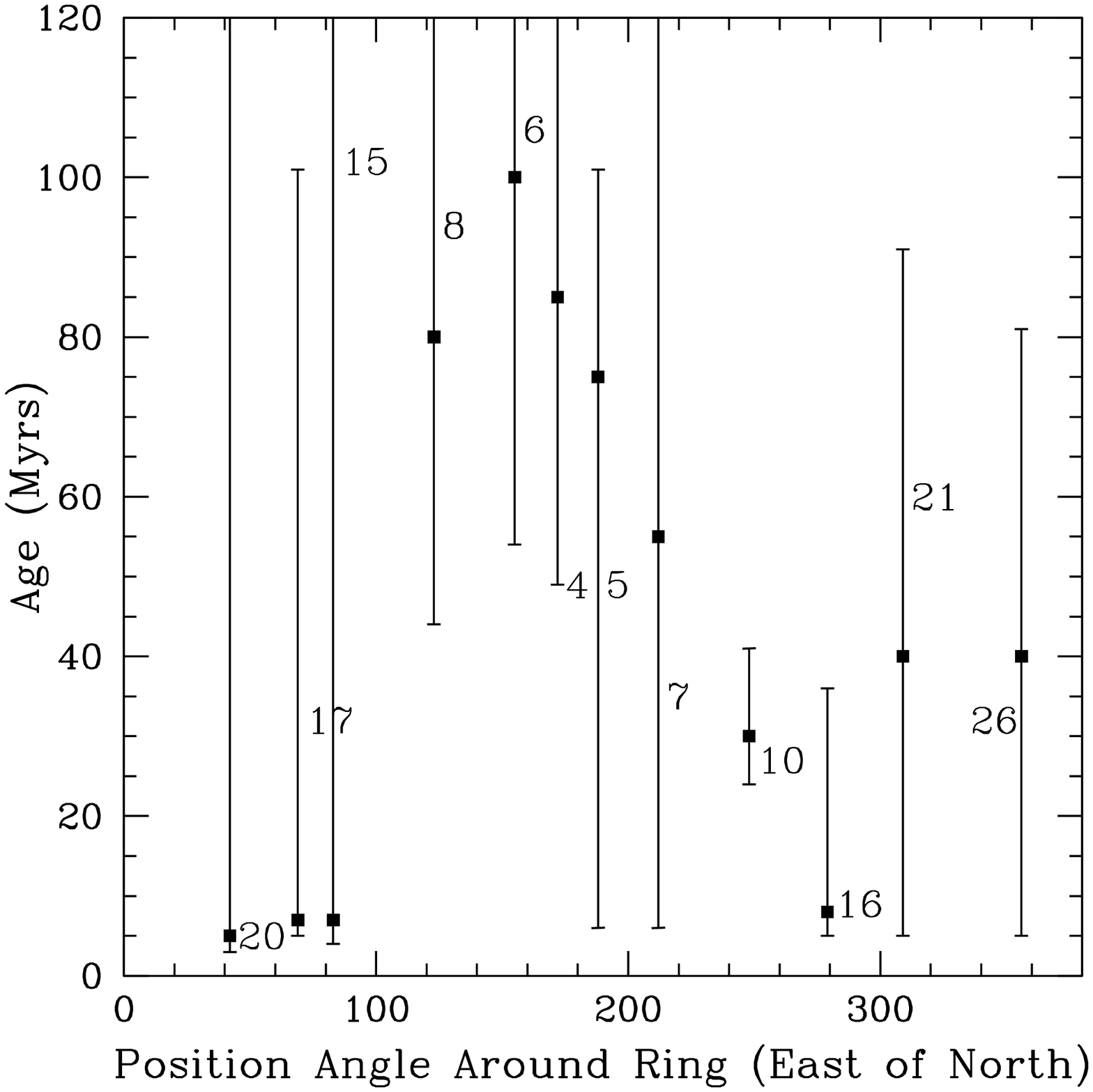}{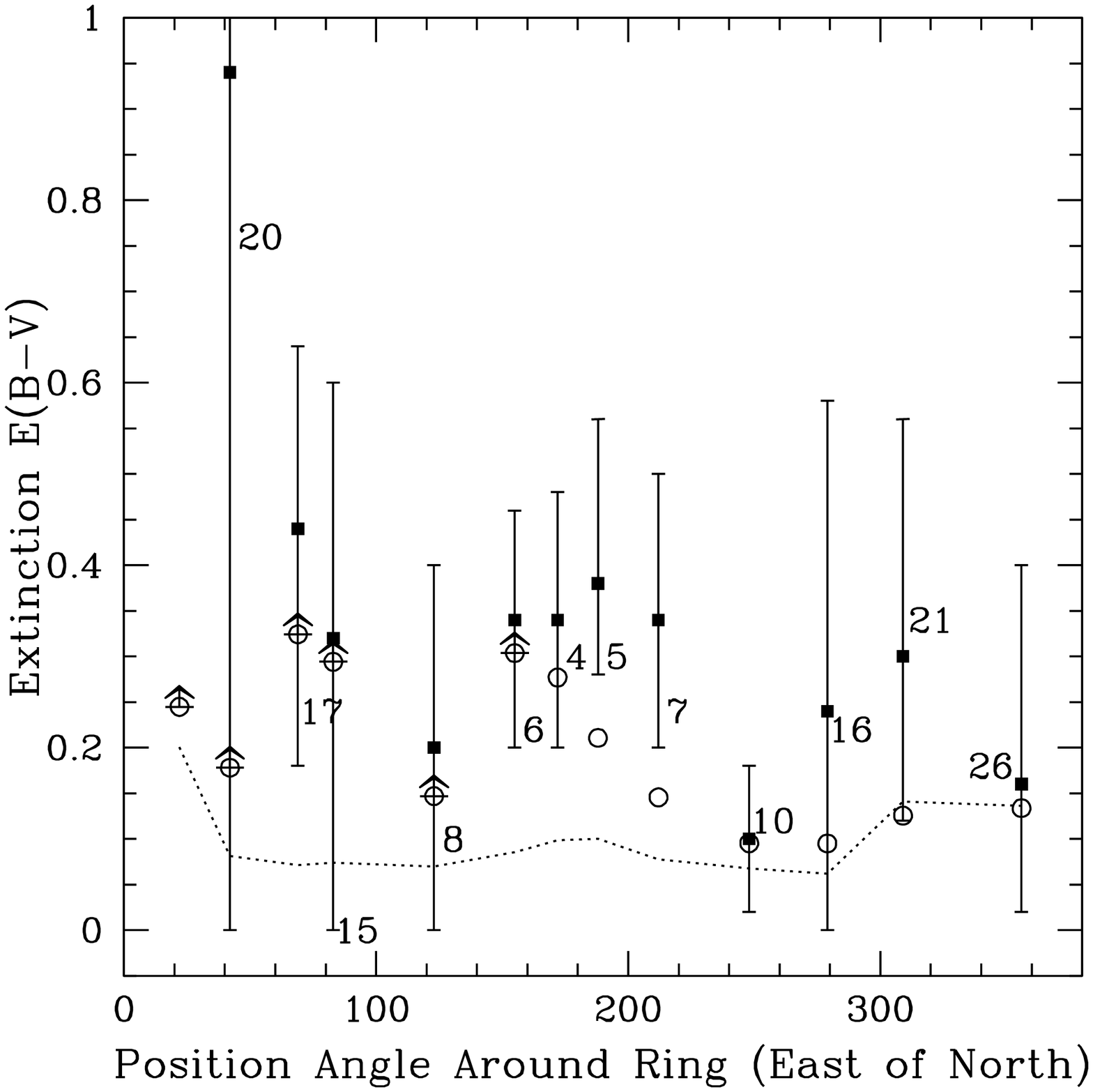}
\caption{Left: For a single-age instantaneous burst model,
a plot of age as a function of position angle (east of north) for the 
clumps along the spiral arm/ring.
The individual clumps are labeled.
Right: A plot of reddening E(B $-$ V) around the ring, as determined
using a single-age instantaneous burst model (filled squares).
The errorbars plotted in both of these figures were deriving by
adding in quadrature the statistical uncertainties in the photometry,
uncertainties in the colors due to background subtraction,
and the uncertainties in the GALEX aperture corrections
(see text for more details).
The dotted curve marks the dust attenuations derived using
the FIR/FUV ratio, as described in the text.
The open circles are reddenings determined from the 
L$_{H\alpha}$/L$_{24{\mu}m}$ ratios (see text).  
Note that the E(B $-$ V) values 
determined from 
L$_{H\alpha}$/L$_{24{\mu}m}$ 
are lower limits for position angles between 0\deg and 160\deg.
}
\label{fig5}
\end{figure}

\begin{figure}
\plottwo{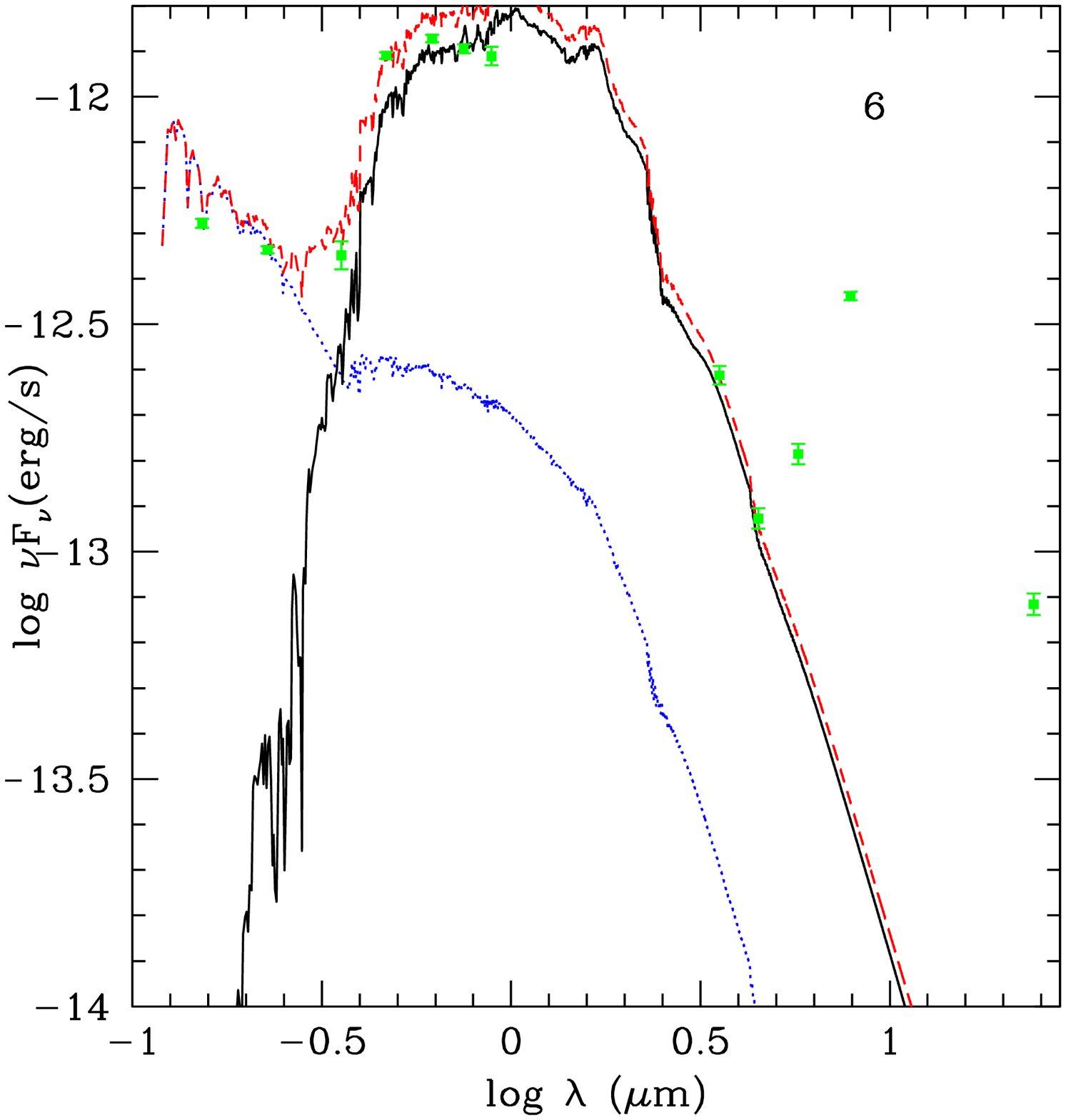}{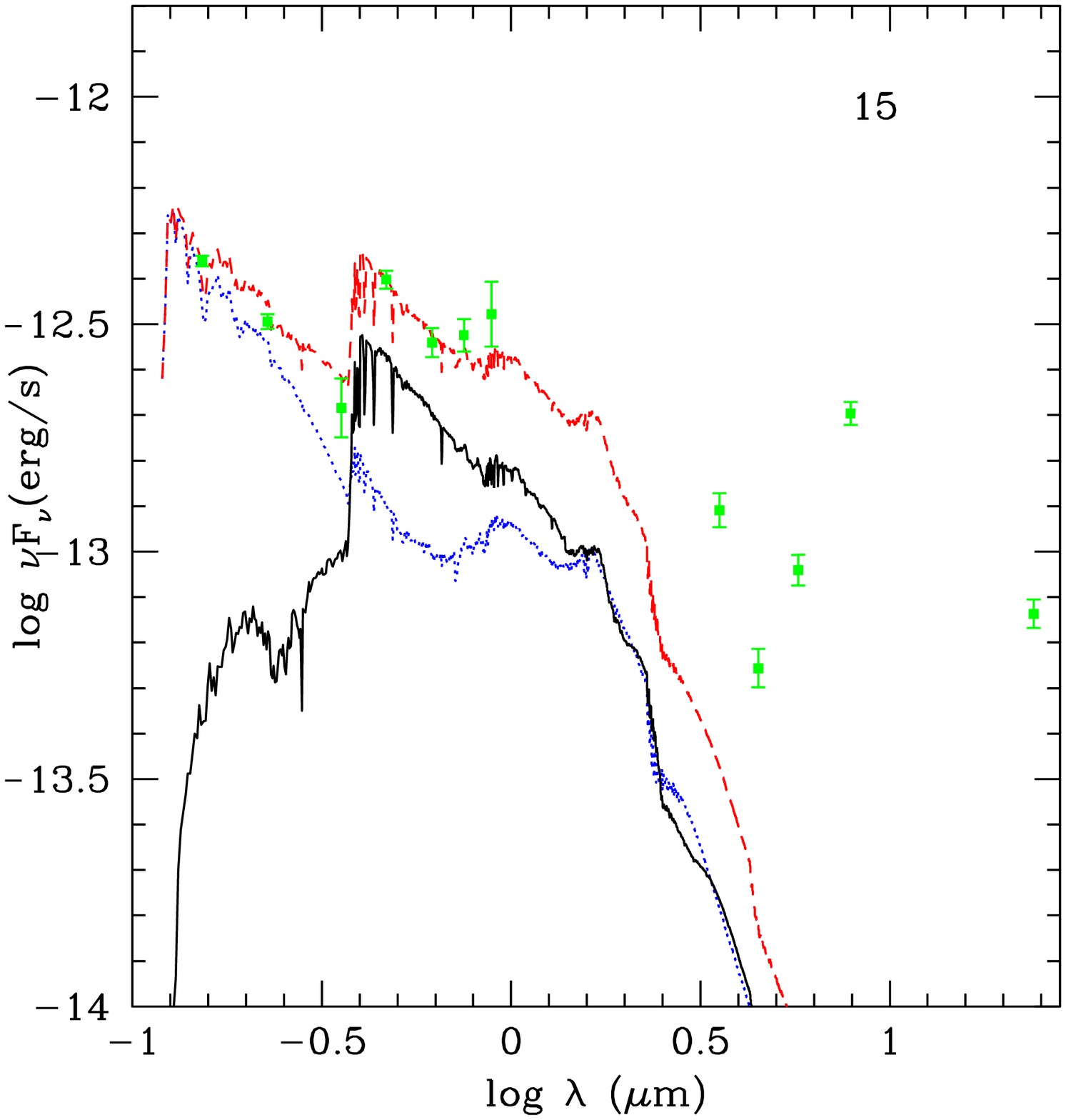}
\caption{Example two-component fits for clumps 6 and 15, as discussed
in Section 8.2.
Only statistical uncertainties are shown.
The solid black curve is the older less-extincted component, the blue dotted curve
is the younger more extincted component, while the red dashed curve is the sum
of the two components.
For clump 6, the younger component is 8 Myrs old, extincted by E(B$-$V) = 0.08.
The older component is 1500 Myrs old, with no attenuation.
The young stars/old stars mass ratio is 0.0061.
For clump 15, the younger population is 15 Myrs old with 
E(B $-$ V) = 0.04, and the older is
350 Myrs with zero attenuation.
The young stars/old stars mass ratio is 0.079.
}
\label{fig6}
\end{figure}

\begin{figure}
\plottwo{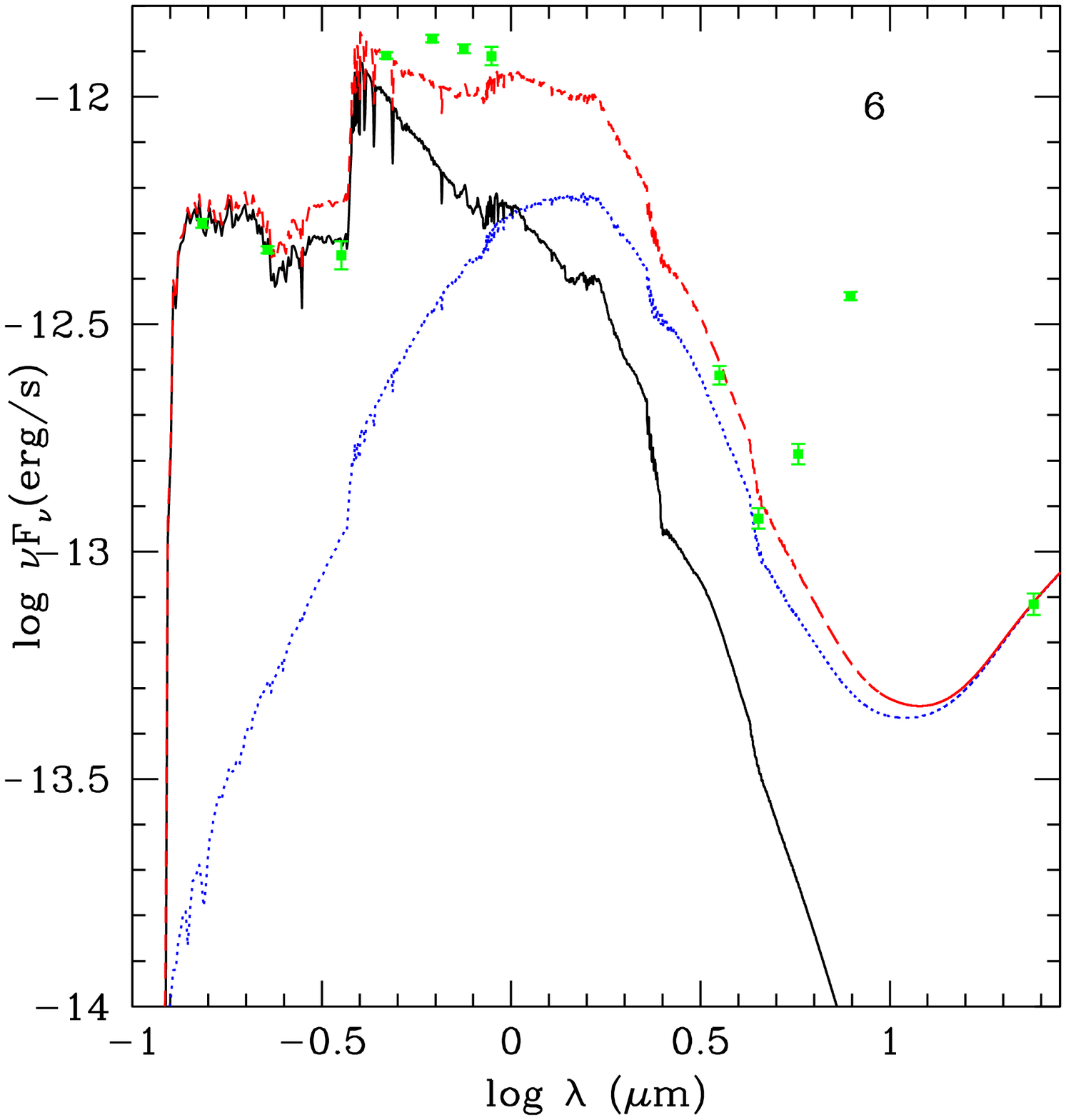}{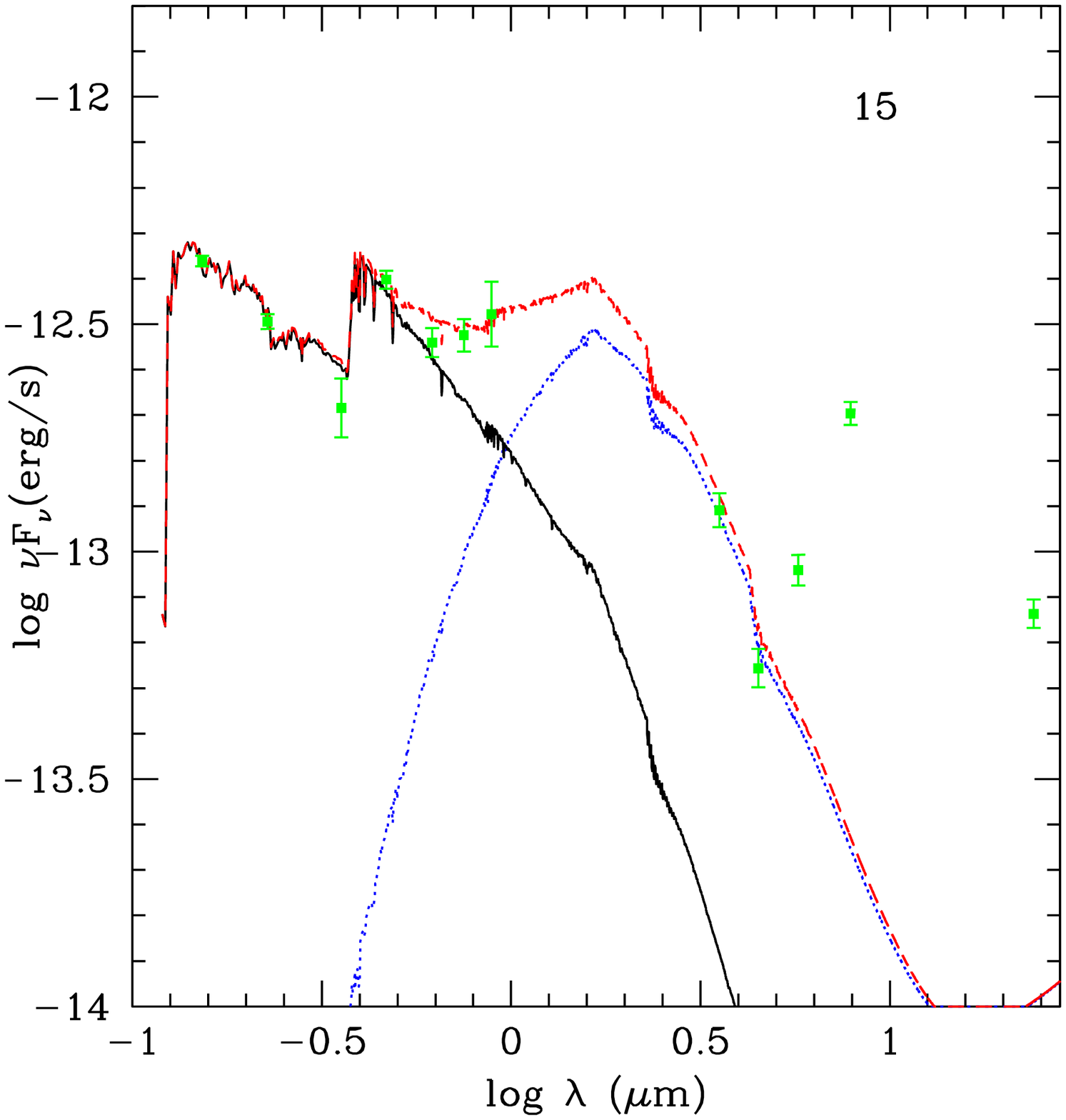}
\caption{An alternative two-component decomposition of the SEDs for clumps 6 and 15,
as discussed in Section 8.3.
Only statistical uncertainties are shown.
The solid black curve is the older less-extincted component, the blue dotted curve
is the younger more extincted component, while the red dashed curve is the sum
of the two components.
For clump 6, the younger component is 6 Myrs old, 
extincted by E(B$-$V) = 0.8.
The older component is 200 Myrs old, with no reddening.
The young stars/old stars mass ratio is 0.79.
For clump 15, the younger population is 8 Myrs old with 
E(B $-$ V) = 1, and the older is
75 Myrs with zero attenuation.
The young stars/old stars mass ratio is 2.4.
}
\label{fig7}
\end{figure}

\begin{figure}
\plottwo{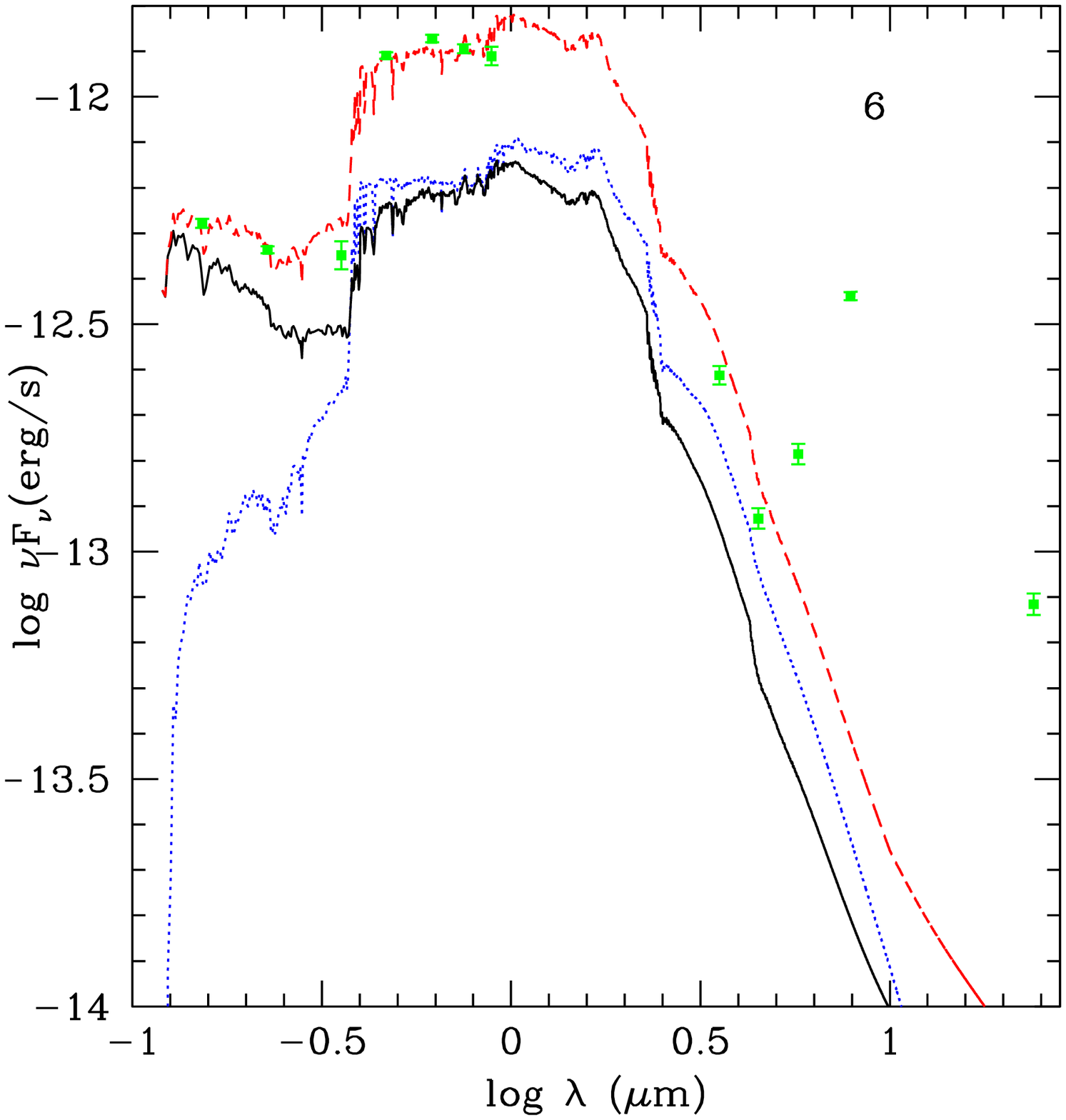}{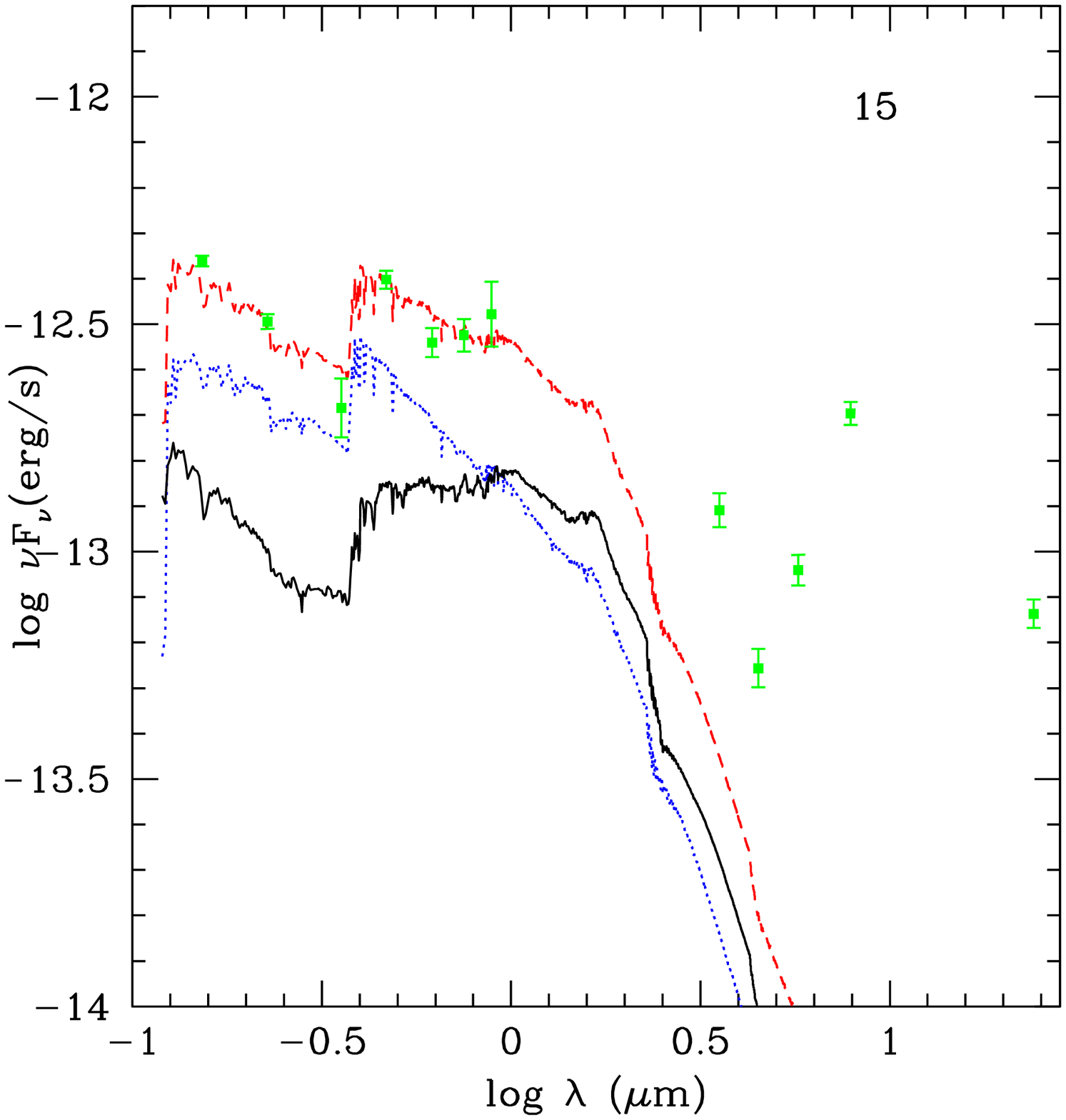}
\caption{An alternative two-component decomposition of the SEDs for clumps 6 and 15,
as discussed in Section 8.4.  These models consist of continuous star formation
starting 10 Gyrs ago combined with a more recent, more
reddened instantaneous burst. 
Only statistical uncertainties are shown.
The solid black curve is the less-extincted component (continuous
star formation), the blue dotted curve
is the younger more extincted component, while the red dashed curve is the sum
of the two components.
For clump 6, the younger component is 200 Myrs old, 
extincted by E(B$-$V) = 0.22.
For clump 15, the younger population is 50 Myrs old with 
E(B $-$ V) = 0.10.
For both decompositions, E(B $-$ V)(continuous star formation) = 0.44 $\times$
E(B $-$ V)(intermediate age).
}
\label{fig8}
\end{figure}

\begin{figure}
\plottwo{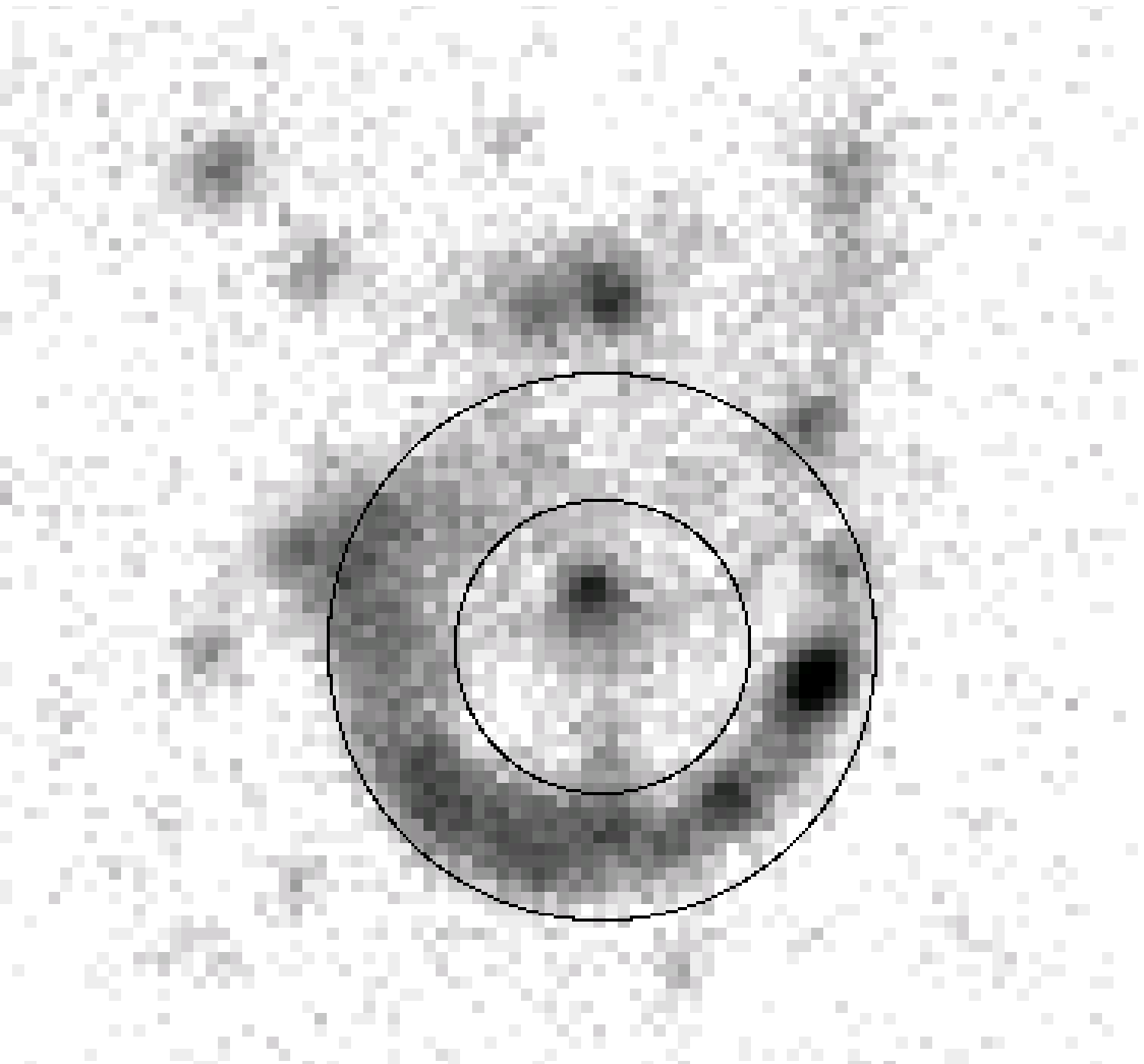}{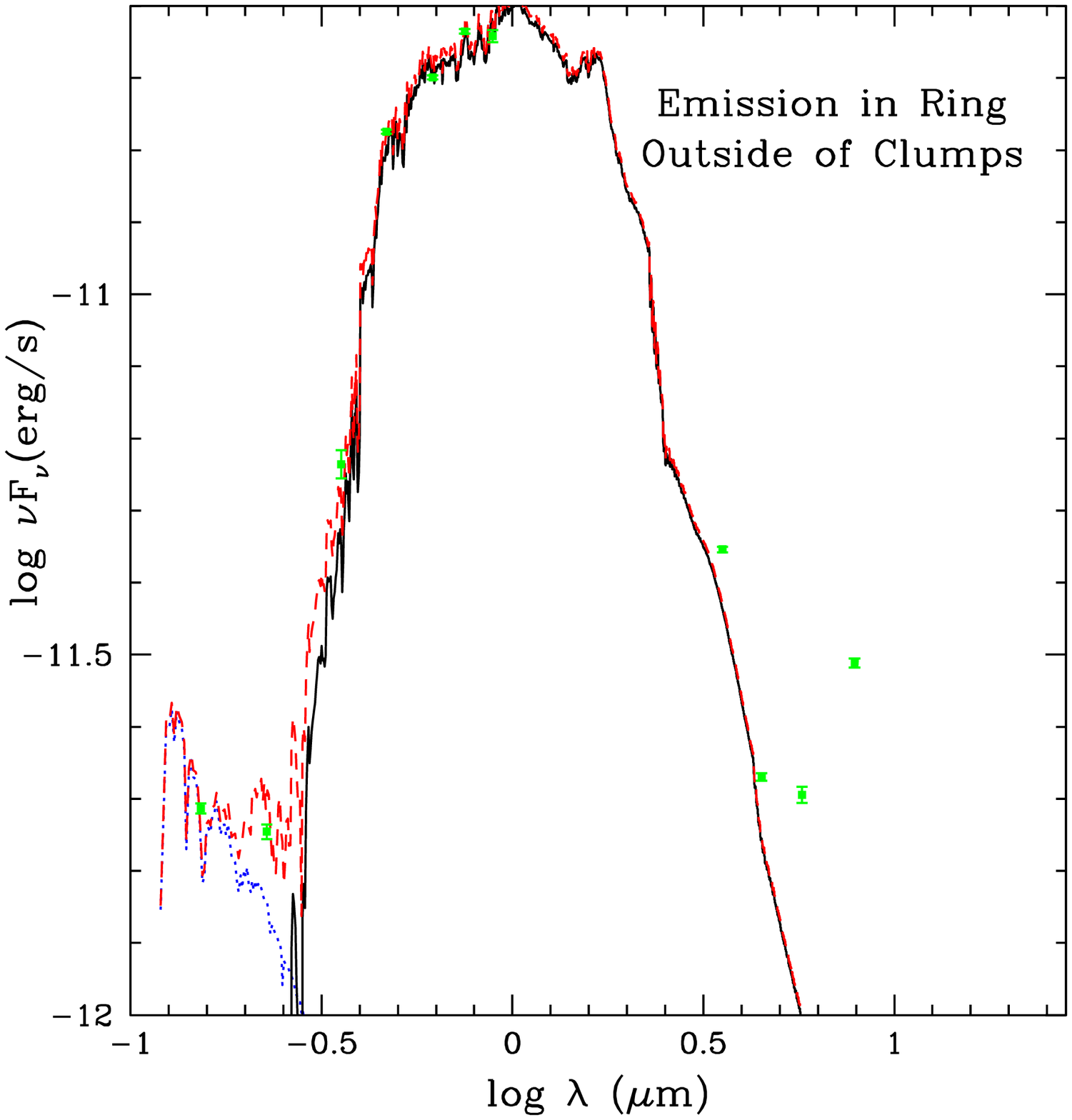}
\caption{Left: The annulus in which the interclump emission was 
extracted, superimposed on the GALEX NUV image (from
\citealp{smi10}).
The field of view and orientation is the same as in Figure 1.
Right: The SED of the interclump emission, after the fluxes of the clumps
were subtracted.
Only statistical uncertainties are shown.
The solid black curve is an unextincted 1500 Myrs old population, 
the blue dotted curve
is an 8 Myrs old population extincted by E(B$-$V) = 0.08,
and the red dashed curve is the sum
of the two components.
The young stars/old stars mass ratio is 0.0011.
}
\label{fig9}
\end{figure}

\begin{figure}
\includegraphics[scale=0.5]{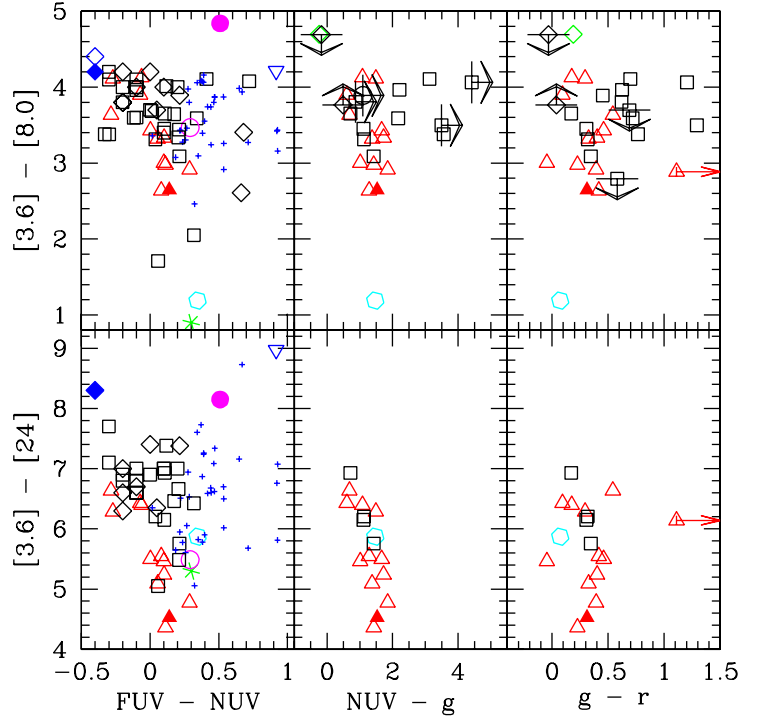}
\caption{Various UV/optical/IR color-color plots 
for star forming clumps
within Arp 107 and other interacting galaxies. 
For Arp 107, only the clumps in the arm/ring (red open triangles)
and the end of the tail (clump 29; filled red triangle) are included.
The magenta filled circle marks Stephan's Quintet \#5,
the open magenta circle is Stephan's Quintet \#2,
and the open cyan hexagon marks 
the VCC 2062 
(from \citealp{boquien10}).
The open blue diamond is the clump at the end of the Arp 82 tail, while
the green asterisk marks clump \#26 in the Arp 82 tail and 
the blue filled diamond is the Arp 82 hinge clump (clump 21 at the base of the Arp 82
tail) 
(from \citealp{hancock07}).
The small blue plus signs are clumps in Arp 244, while the
upside down open blue triangle is the `hot spot' in the `overlap
region' in Arp 244 (from 
\citealp{zhang10}).
The green open diamond is the IR-bright clump in the northern Arp 285 tail
\citep{smi08}.
The black symbols mark the locations of additional clumps from these galaxies
and others, including Arp 24 \citep{cao07}, Arp 284 \citep{peterson09},
and Arp 143 \citep{beirao09},
along with Arp 105, Arp 245, NGC 5291, and NGC 7252 \citep{boquien10}.
The open black squares are disk clumps and the open
black diamonds are tail clumps.  
For clarity, errorbars are omitted.   These are generally about the size of
the data points or slightly larger.
}\label{fig10}
\end{figure}

\begin{figure}
\includegraphics[scale=0.5]{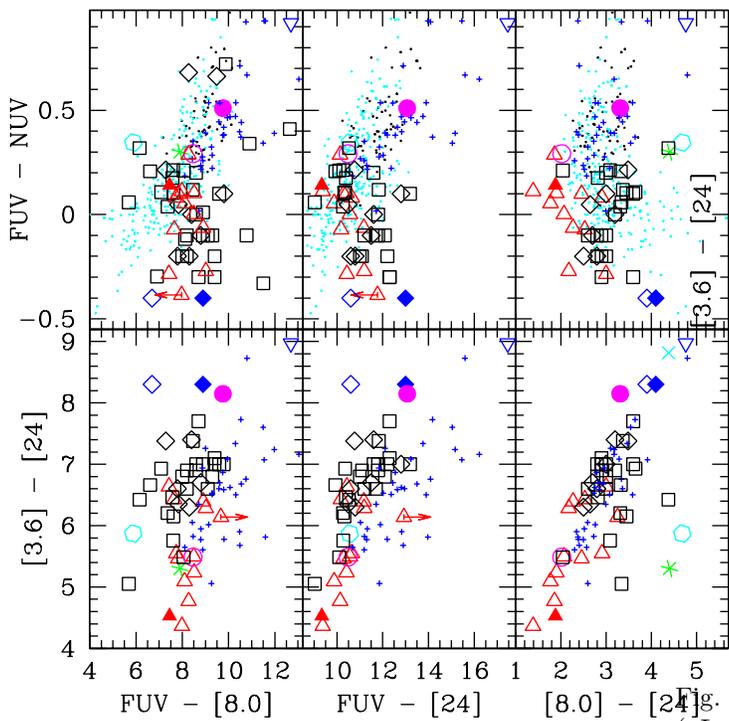}
\caption{Additional UV/optical/IR color-color plots for star forming clumps
within Arp 107 and other interacting galaxies. 
The symbols are the same as in Figure 10, with the addition 
of clump i from NGC 2207 (the cyan cross, from \citealp{elmegreen06})
and H~II regions from nearby normal spiral galaxies (cyan dots)
and from Arp 85 (M51) (black dots) (from \citealp{boquien09}).
}\label{fig11}
\end{figure}

\begin{figure}
\includegraphics[scale=0.5]{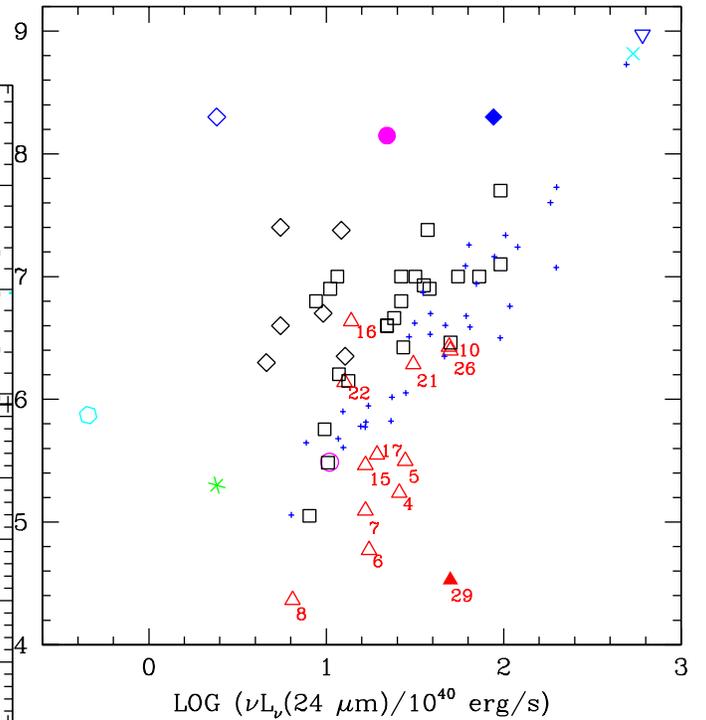}
\caption{A comparison of the 24 $\mu$m luminosities ($\nu$L$_{\nu}$)
and the [3.6] $-$ [24] colors of the Arp 107 clumps and clumps
in other interacting galaxies.
The symbols are the same as in Figures 10 and 11:
Arp 107 arm/ring (red open triangle), 
Arp 107 end of tail (red filled triangle), 
SQ 5 (magenta filled circle), 
Arp 82 hinge clump (blue filled diamond),
Arp 82 clump 26 (green asterisk), VCC 2062 (open cyan hexagon),
NGC 2207 clump i (cyan cross),
Arp 244 hot spot (blue upside down triangle),
additional regions in Arp 244 (small plus signs), 
other disk clumps (open black squares),
and 
other tail clumps (open black diamonds).
The Arp 107 clumps are labeled in red.
}\label{fig12}
\end{figure}

\begin{figure}
\plotone{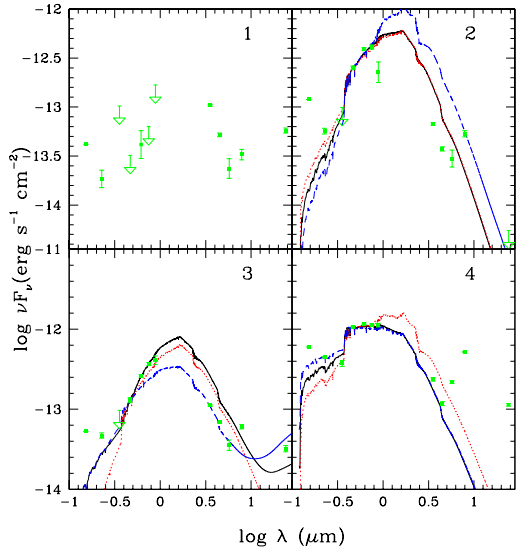}
\caption{SED plots of clumps 1 $-$ 4 (green filled squares).  
The black solid curve is the best-fit single-age instantaneous burst model.
The blue dashed curve is the burst model with the
best-fit reddening, and an age 1$\sigma$ less than 
the best fit age.
The red dotted curve is the burst model with the
best-fit reddening, and an age 1$\sigma$ more than 
the best fit age.
The models have been scaled to match the observed SDSS g band flux.
The plotted error bars only include statistical errors.
}
\end{figure}

\begin{figure}
\plotone{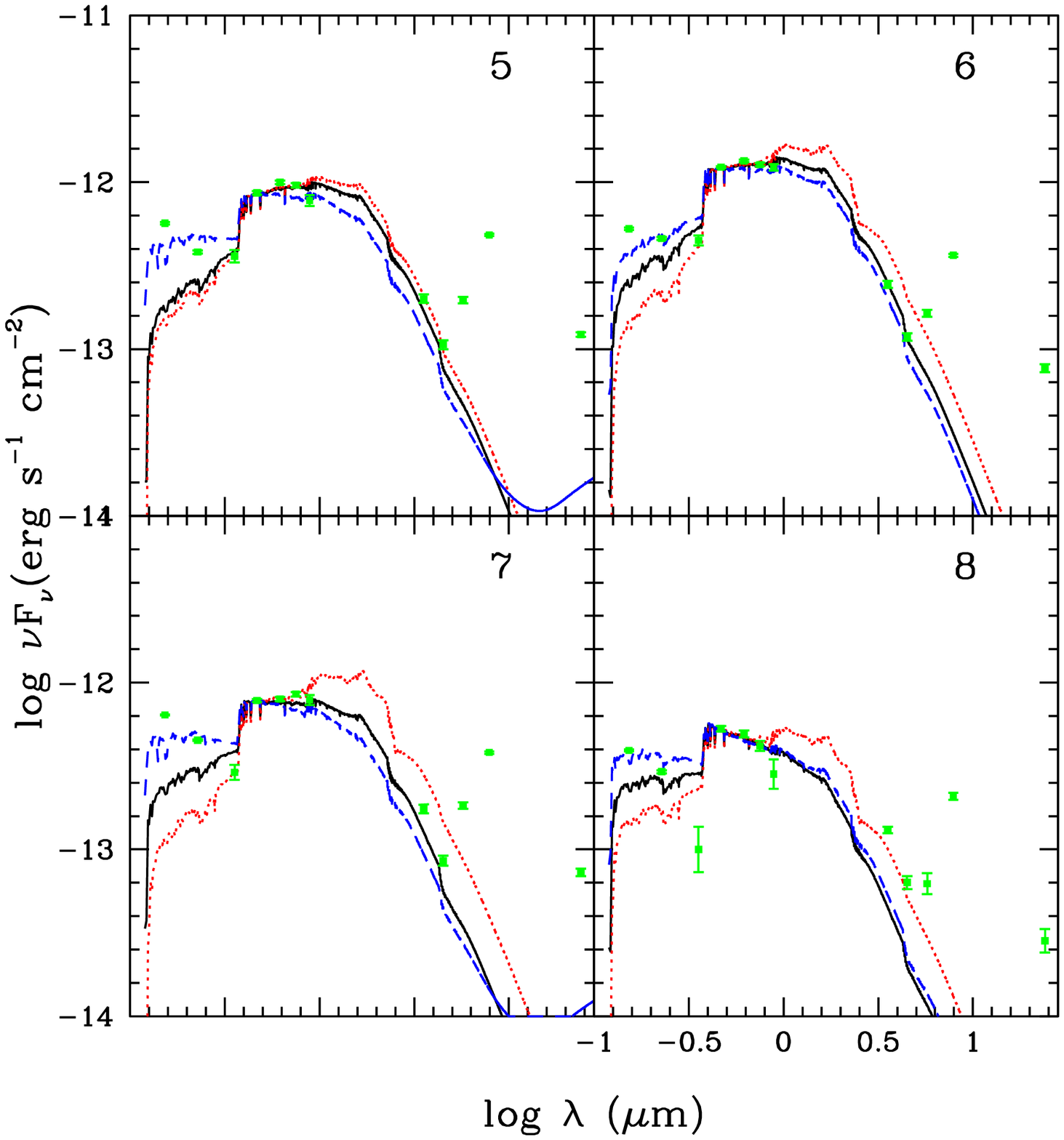}
\caption{SED plots of clumps 5 $-$ 8.  
The black solid curve is the best-fit single-age instantaneous burst model.
The blue dashed curve is the burst model with the
best-fit reddening, and an age 1$\sigma$ less than 
the best fit age.
The red dotted curve is the burst model with the
best-fit reddening, and an age 1$\sigma$ more than 
the best fit age.
The plotted error bars only include statistical errors.
}
\end{figure}

\begin{figure}
\plotone{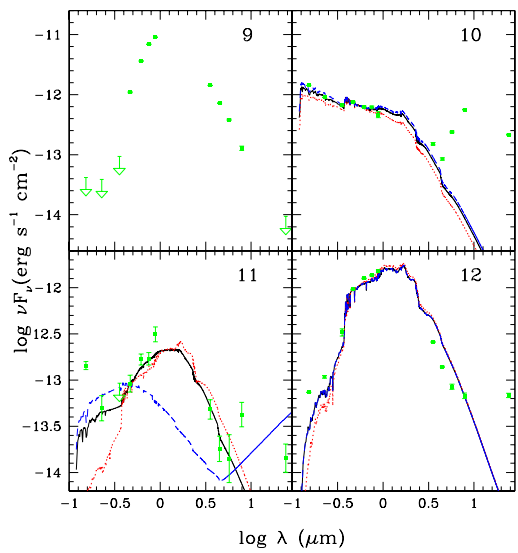}
\caption{SED plots of clumps 9 $-$ 12.  
The black solid curve is the best-fit single-age instantaneous burst model.
The blue dashed curve is the burst model with the
best-fit reddening, and an age 1$\sigma$ less than 
the best fit age.
The red dotted curve is the burst model with the
best-fit reddening, and an age 1$\sigma$ more than 
the best fit age.
Note that clump 9 is a foreground star (see text).
The plotted error bars only include statistical errors.
}
\end{figure}

\begin{figure}
\plotone{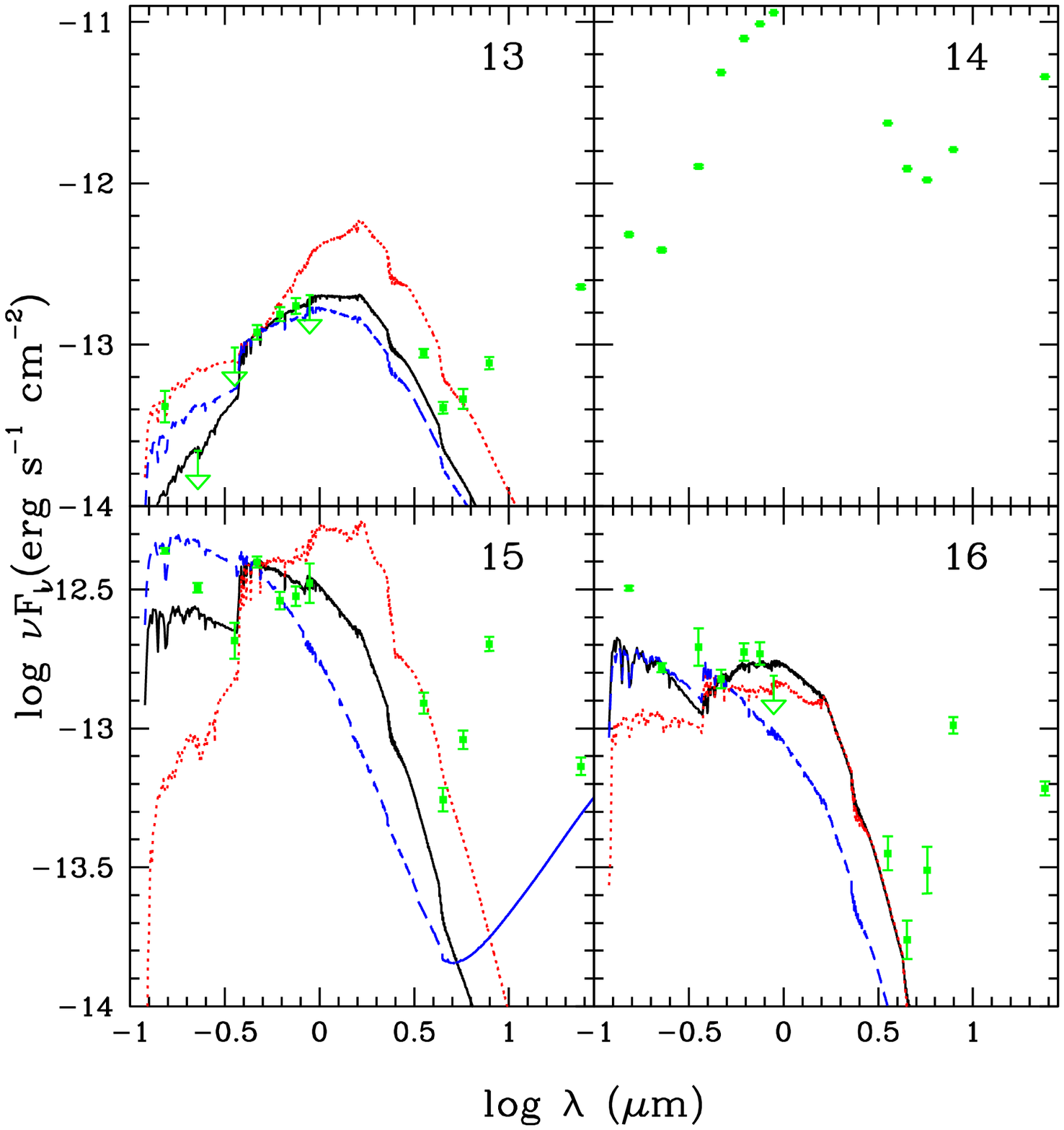}
\caption{SED plots of clumps 13 $-$ 16.  
The black solid curve is the best-fit single-age instantaneous burst model.
The blue dashed curve is the burst model with the
best-fit reddening, and an age 1$\sigma$ less than 
the best-fit age.
The red dotted curve is the burst model with the
best-fit reddening, and an age 1$\sigma$ more than 
the best-fit age.
The plotted error bars only include statistical errors.
}
\end{figure}

\begin{figure}
\plotone{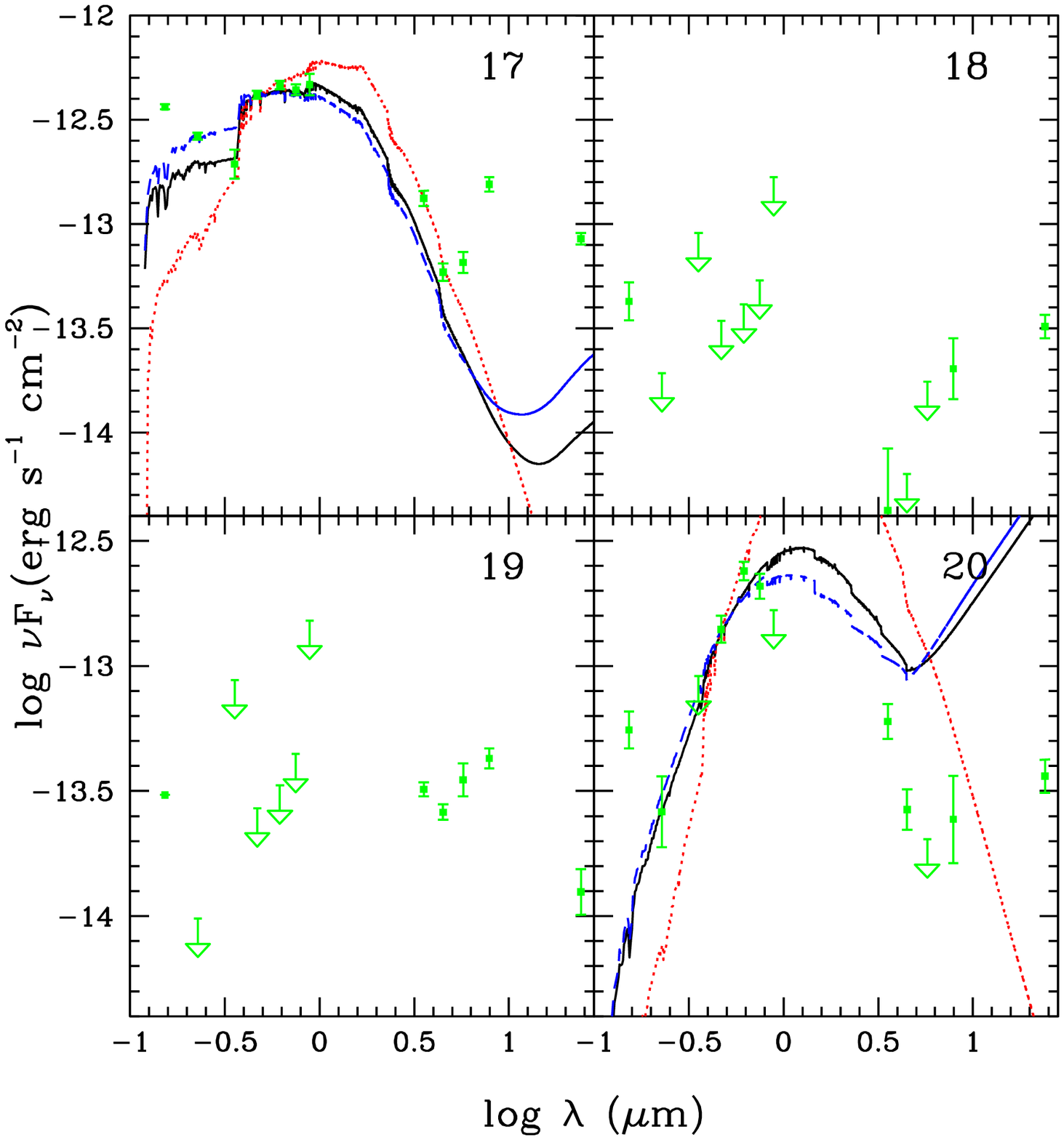}
\caption{SED plots of clumps 17 $-$ 20.  
The black solid curve is the best-fit single-age instantaneous burst model.
The blue dashed curve is the burst model with the
best-fit reddening, and an age 1$\sigma$ less than 
the best fit age.
The red dotted curve is the burst model with the
best-fit reddening, and an age 1$\sigma$ more than 
the best fit age.
The plotted error bars only include statistical errors.
}
\end{figure}

\clearpage

\begin{figure}
\plotone{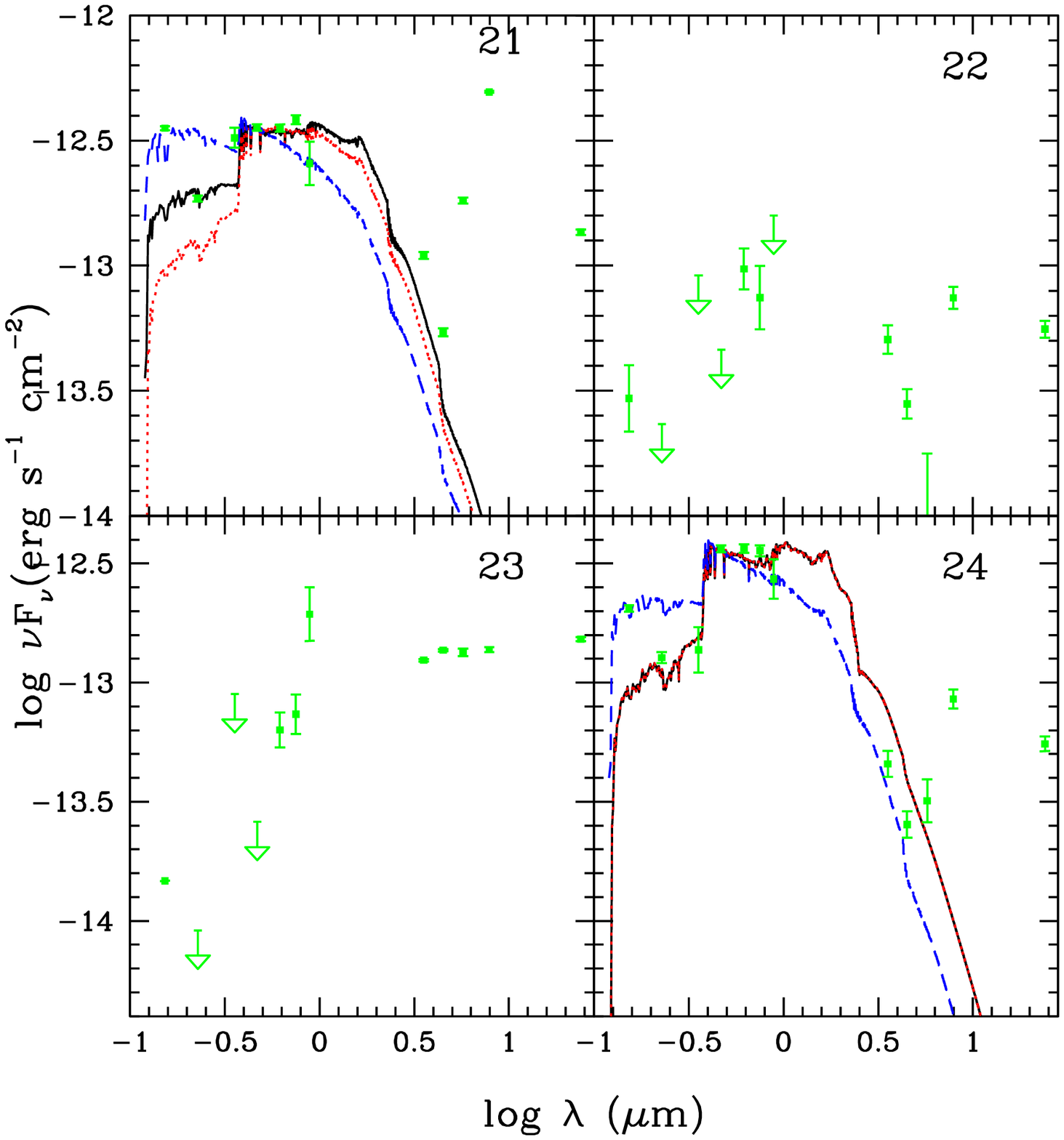}
\caption{SED plots of clumps 21 $-$ 24.  
The black solid curve is the best-fit single-age instantaneous burst model.
The blue dashed curve is the burst model with the
best-fit reddening, and an age 1$\sigma$ less than 
the best fit age.
The red dotted curve is the burst model with the
best-fit reddening, and an age 1$\sigma$ more than 
the best fit age.
The plotted error bars only include statistical errors.
}
\end{figure}

\begin{figure}
\plotone{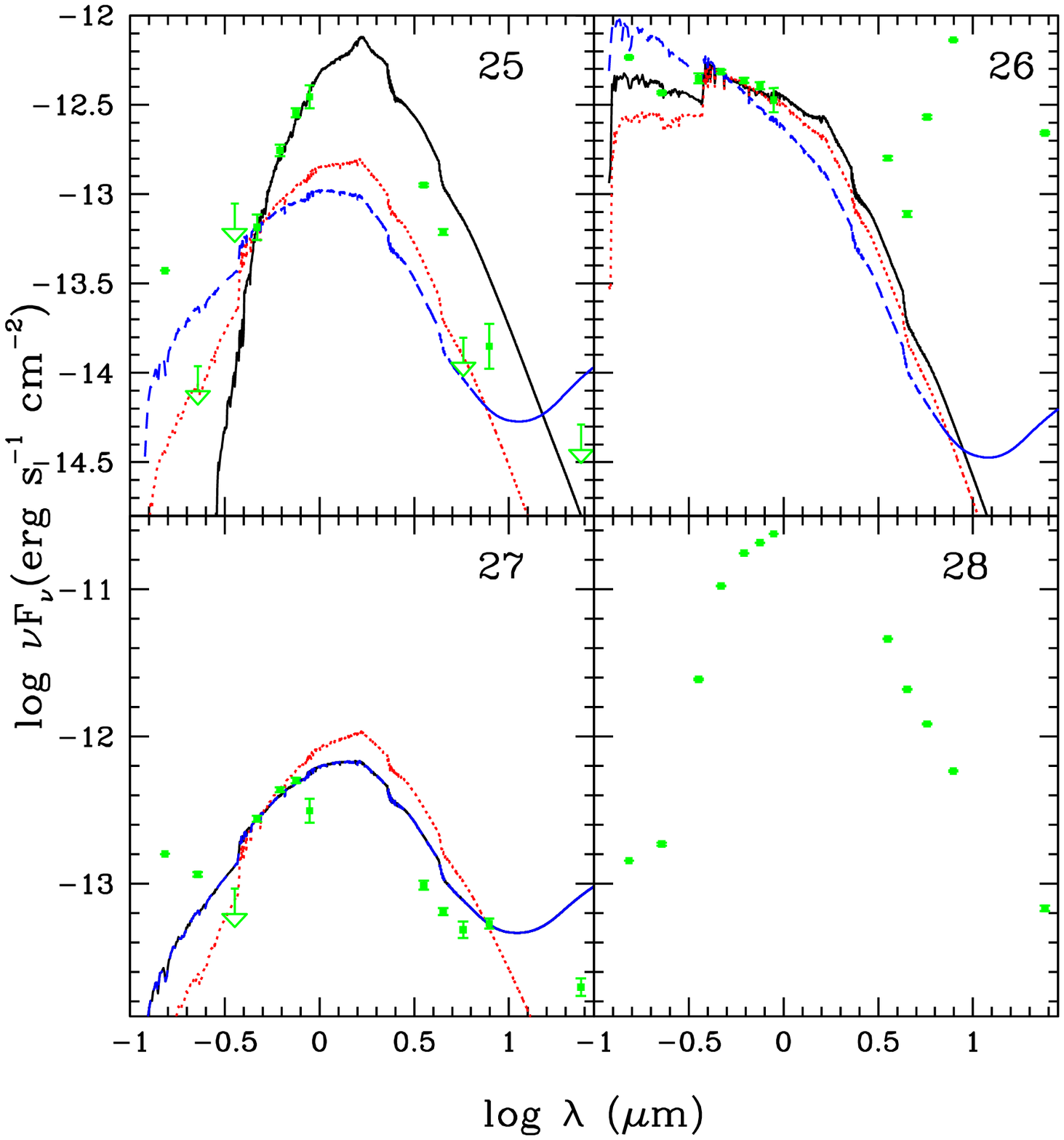}
\caption{SED plots of clumps 25 $-$ 28.  
The black solid curve is the best-fit single-age instantaneous burst model.
The blue dashed curve is the burst model with the
best-fit reddening, and an age 1$\sigma$ less than 
the best fit age.
The red dotted curve is the burst model with the
best-fit reddening, and an age 1$\sigma$ more than 
the best fit age.
The plotted error bars only include statistical errors.
}
\end{figure}

\begin{figure}
\plotone{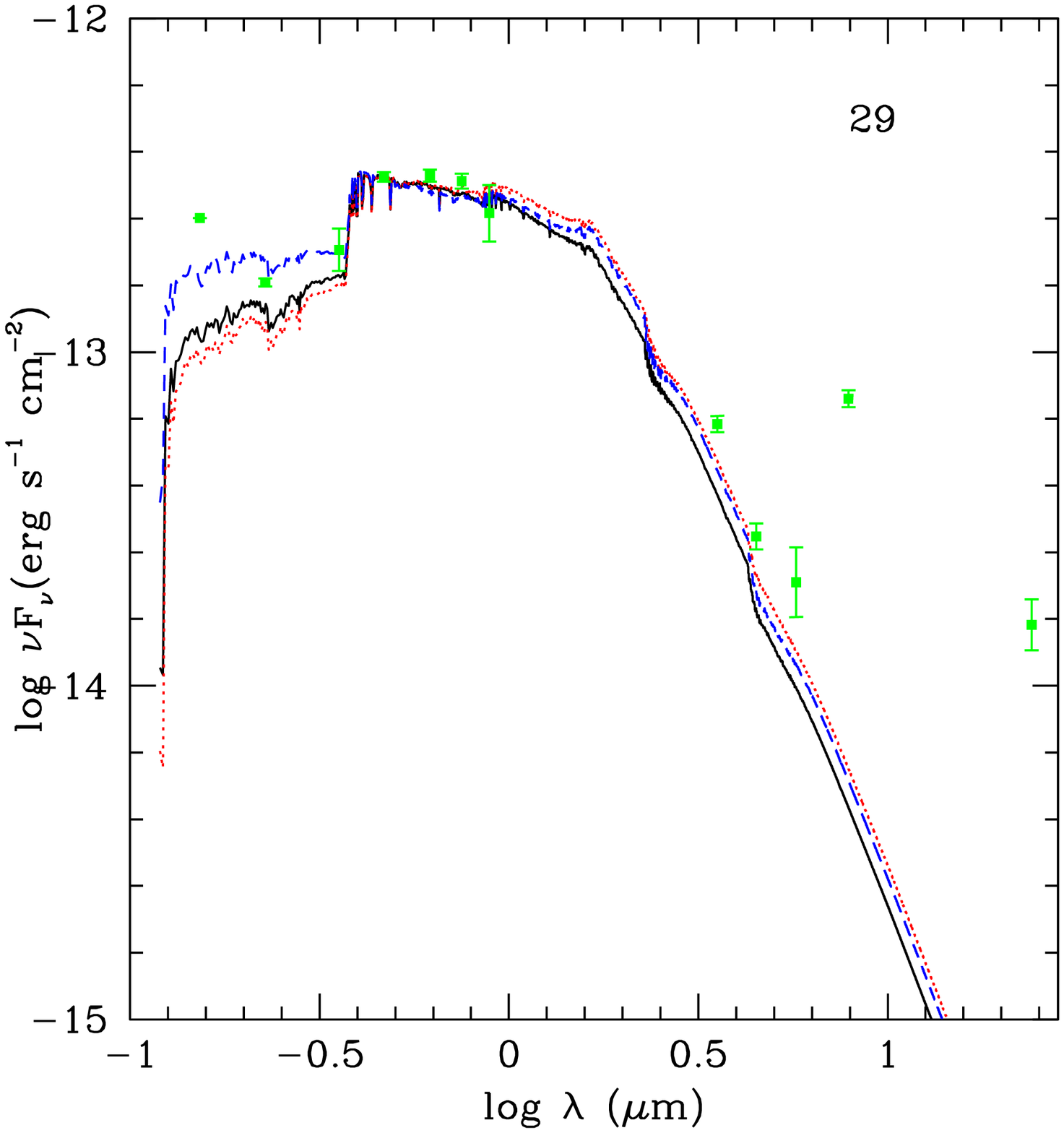}
\caption{SED plot of clump 29.  
The black solid curve is the best-fit single-age instantaneous burst model.
The blue dashed curve is the burst model with the
best-fit reddening, and an age 1$\sigma$ less than 
the best fit age.
The red dotted curve is the burst model with the
best-fit reddening, and an age 1$\sigma$ more than 
the best fit age.
The plotted error bars only include statistical errors.
}
\end{figure}

\begin{deluxetable}{cccccccc}
\tabletypesize{\scriptsize}
\setlength{\tabcolsep}{0.03in}
\def\et#1#2#3{${#1}^{+#2}_{-#3}$}
\tablewidth{0pt}
\tablecaption{UV and Optical Magnitudes for Clumps in Arp 107\label{tab-2}}
\tablehead{
\multicolumn{1}{c}{ID} &
\multicolumn{1}{c}{FUV} &
\multicolumn{1}{c}{NUV} &
\multicolumn{1}{c}{u} &
\multicolumn{1}{c}{g} &
\multicolumn{1}{c}{r} &
\multicolumn{1}{c}{i} &
\multicolumn{1}{c}{z} 
\\ 
\multicolumn{1}{c}{} &
\multicolumn{1}{c}{(mag)} &
\multicolumn{1}{c}{(mag)} &
\multicolumn{1}{c}{(mag)} &
\multicolumn{1}{c}{(mag)} &
\multicolumn{1}{c}{(mag)} &
\multicolumn{1}{c}{(mag)} &
\multicolumn{1}{c}{(mag)} 
\\ 
}
\startdata

  1   &    22.89 $\pm$  0.01   &    23.33 $\pm$  0.22   & $\ge$21.06          & $\ge$22.06          &    21.51 $\pm$  0.35   & $\ge$20.86           &  $\ge$19.62          \\
  2   &    21.74 $\pm$  0.01   &    22.11 $\pm$  0.07   & $\ge$21.10          &    19.82 $\pm$  0.05   &    19.07 $\pm$  0.04   &    18.83 $\pm$  0.05    &     19.30 $\pm$  0.26   \\
  3   &    22.63 $\pm$  0.01   &    22.32 $\pm$  0.08   & $\ge$21.13          &    20.54 $\pm$  0.08   &    19.51 $\pm$  0.05   &    18.94 $\pm$  0.05    &     18.65 $\pm$  0.13   \\
  4   &    19.98 $\pm$  0.02   &    19.88 $\pm$  0.02   &    19.66 $\pm$  0.09   &    18.25 $\pm$  0.02   &    17.88 $\pm$  0.02   &    17.72 $\pm$  0.03    &     17.56 $\pm$  0.06   \\
  5   &    20.05 $\pm$  0.02   &    20.05 $\pm$  0.03   &    19.70 $\pm$  0.10   &    18.48 $\pm$  0.02   &    18.05 $\pm$  0.03   &    17.90 $\pm$  0.03    &     17.96 $\pm$  0.08   \\
  6   &    20.13 $\pm$  0.02   &    19.84 $\pm$  0.02   &    19.46 $\pm$  0.08   &    18.10 $\pm$  0.02   &    17.73 $\pm$  0.02   &    17.59 $\pm$  0.03    &     17.46 $\pm$  0.05   \\
  7   &    19.92 $\pm$  0.01   &    19.86 $\pm$  0.01   &    19.94 $\pm$  0.11   &    18.59 $\pm$  0.02   &    18.29 $\pm$  0.03   &    18.03 $\pm$  0.03    &     17.95 $\pm$  0.08   \\
  8   &    20.45 $\pm$  0.02   &    20.34 $\pm$  0.02   &    21.09 $\pm$  0.34   &    19.02 $\pm$  0.04   &    18.82 $\pm$  0.05   &    18.81 $\pm$  0.08    &     19.06 $\pm$  0.22   \\
  9   & $\ge$22.88          & $\ge$22.53          & $\ge$21.16          &    18.22 $\pm$  0.02   &    16.65 $\pm$  0.01   &    15.76 $\pm$  0.00    &     15.29 $\pm$  0.01   \\
 10   &    19.05 $\pm$  0.01   &    19.12 $\pm$  0.01   &    19.04 $\pm$  0.05   &    18.62 $\pm$  0.02   &    18.56 $\pm$  0.03   &    18.37 $\pm$  0.04    &     18.52 $\pm$  0.13   \\
 11   &    21.54 $\pm$  0.11   &    22.26 $\pm$  0.34   & $\ge$21.18          &    20.92 $\pm$  0.23   &    19.99 $\pm$  0.15   &    19.78 $\pm$  0.17    &     18.95 $\pm$  0.20   \\
 12   &    22.27 $\pm$  0.01   &    21.41 $\pm$  0.04   &    19.80 $\pm$  0.10   &    18.37 $\pm$  0.01   &    17.80 $\pm$  0.01   &    17.52 $\pm$  0.01    &     17.24 $\pm$  0.04   \\
 13   &    22.89 $\pm$  0.24   & $\ge$23.14          & $\ge$21.13          &    20.63 $\pm$  0.11   &    20.07 $\pm$  0.11   &    19.76 $\pm$  0.12    &  $\ge$19.42          \\
 14   &    20.23 $\pm$  0.02   &    20.03 $\pm$  0.02   &    18.33 $\pm$  0.03   &    16.61 $\pm$  0.00   &    15.81 $\pm$  0.00   &    15.39 $\pm$  0.00    &     15.04 $\pm$  0.01   \\
 15   &    20.34 $\pm$  0.03   &    20.24 $\pm$  0.04   &    20.30 $\pm$  0.16   &    19.33 $\pm$  0.05   &    19.40 $\pm$  0.08   &    19.17 $\pm$  0.09    &     18.88 $\pm$  0.18   \\
 16   &    20.67 $\pm$  0.02   &    20.96 $\pm$  0.04   &    20.36 $\pm$  0.17   &    20.38 $\pm$  0.09   &    19.86 $\pm$  0.08   &    19.69 $\pm$  0.10    &  $\ge$19.71          \\
 17   &    20.53 $\pm$  0.03   &    20.45 $\pm$  0.04   &    20.38 $\pm$  0.17   &    19.27 $\pm$  0.05   &    18.88 $\pm$  0.05   &    18.75 $\pm$  0.06    &     18.52 $\pm$  0.13   \\
 18   &    22.86 $\pm$  0.22   & $\ge$23.29          & $\ge$21.20          & $\ge$21.99          & $\ge$21.51          & $\ge$21.03           &  $\ge$19.63          \\
 19   &    23.22 $\pm$  0.01   & $\ge$24.03          & $\ge$21.24          & $\ge$22.24          & $\ge$21.74          & $\ge$21.24           &  $\ge$19.74          \\
 20   &    22.57 $\pm$  0.19   &    22.96 $\pm$  0.35   & $\ge$21.20          &    20.46 $\pm$  0.13   &    19.60 $\pm$  0.09   &    19.56 $\pm$  0.13    &  $\ge$19.63          \\
 21   &    20.56 $\pm$  0.02   &    20.83 $\pm$  0.03   &    19.82 $\pm$  0.10   &    19.44 $\pm$  0.03   &    19.18 $\pm$  0.04   &    18.90 $\pm$  0.04    &     19.17 $\pm$  0.22   \\
 22   &    23.26 $\pm$  0.33   & $\ge$23.09          & $\ge$21.19          & $\ge$21.67          &    20.58 $\pm$  0.21   &    20.68 $\pm$  0.32    &  $\ge$19.69          \\
 23   &    24.02 $\pm$  0.01   & $\ge$24.10          & $\ge$21.21          & $\ge$22.28          &    21.05 $\pm$  0.18   &    20.69 $\pm$  0.21    &     19.47 $\pm$  0.28   \\
 24   &    21.16 $\pm$  0.03   &    21.24 $\pm$  0.06   &    20.75 $\pm$  0.24   &    19.42 $\pm$  0.04   &    19.14 $\pm$  0.05   &    18.97 $\pm$  0.06    &     19.10 $\pm$  0.21   \\
 25   &    23.00 $\pm$  0.01   & $\ge$23.91          & $\ge$21.23          &    21.28 $\pm$  0.18   &    19.94 $\pm$  0.08   &    19.22 $\pm$  0.06    &     18.83 $\pm$  0.16   \\
 26   &    20.02 $\pm$  0.01   &    20.08 $\pm$  0.02   &    19.47 $\pm$  0.07   &    19.11 $\pm$  0.03   &    18.96 $\pm$  0.04   &    18.84 $\pm$  0.05    &     18.87 $\pm$  0.17   \\
 27   &    21.43 $\pm$  0.01   &    21.34 $\pm$  0.04   & $\ge$21.18          &    19.72 $\pm$  0.05   &    18.95 $\pm$  0.04   &    18.60 $\pm$  0.04    &     18.95 $\pm$  0.20   \\
 28   &    21.54 $\pm$  0.01   &    20.83 $\pm$  0.02   &    17.62 $\pm$  0.01   &    15.76 $\pm$  0.00   &    14.94 $\pm$  0.00   &    14.56 $\pm$  0.00    &     14.24 $\pm$  0.01   \\
 29   &    20.93 $\pm$  0.01   &    20.98 $\pm$  0.03   &    20.33 $\pm$  0.16   &    19.51 $\pm$  0.04   &    19.23 $\pm$  0.04   &    19.08 $\pm$  0.06    &     19.15 $\pm$  0.21   \\
\enddata
\end{deluxetable}

\begin{deluxetable}{cccrc}
\tabletypesize{\scriptsize}
\tablecaption{Population Synthesis Results: Single-Age Instantaneous Burst Models}
\tablewidth{0pt}
\tablehead{
\colhead{ID} & \colhead{Age} & \colhead{E(B-V)} & \colhead{Reduced} & \colhead{Colors Used}
\\
 & \colhead{(Myr)} & \colhead{(mag)} & \colhead{{Chi Squared} } & \colhead{}
\\
 & \colhead{} & \colhead{} & \colhead{{($\chi$$^2$/(N$-$2))} } & \colhead{}
}
\startdata
    2   & 85  $\pm$ $^{ 66 }_{  36 }$ &  0.64  $\pm$ $^{ 0.18 }_{  0.16 }$ &   7.1  &                       FUV-NUV, NUV-g, g-r, r-i, i-z    \\
    3   &  8  $\pm$ $^{ 93 }_{   3 }$ &  0.80  $\pm$ $^{ 0.44 }_{  0.30 }$ &   0.5  &                       FUV-NUV, NUV-g, g-r, r-i, i-z    \\
 **4   & 85  $\pm$ $^{ 66 }_{  36 }$ &  0.34  $\pm$ $^{ 0.14 }_{  0.14 }$ &   2.2  &                  FUV-NUV, NUV-g, u-g, g-r, r-i, i-z    \\
 **5   & 75  $\pm$ $^{ 26 }_{  69 }$ &  0.38  $\pm$ $^{ 0.18 }_{  0.10 }$ &   2.4  &                  FUV-NUV, NUV-g, u-g, g-r, r-i, i-z    \\
 **6   & 100  $\pm$ $^{ 101 }_{  46 }$ &  0.34  $\pm$ $^{ 0.12 }_{  0.14 }$ &   2.9  &                  FUV-NUV, NUV-g, u-g, g-r, r-i, i-z    \\
 **7   & 55  $\pm$ $^{ 96 }_{  49 }$ &  0.34  $\pm$ $^{ 0.16 }_{  0.14 }$ &   2.1  &                  FUV-NUV, NUV-g, u-g, g-r, r-i, i-z    \\
 **8   & 80  $\pm$ $^{ 71 }_{  36 }$ &  0.20  $\pm$ $^{ 0.20 }_{  0.20 }$ &   2.2  &                  FUV-NUV, NUV-g, u-g, g-r, r-i, i-z    \\
 **10   & 30  $\pm$ $^{ 11 }_{   6 }$ &  0.10  $\pm$ $^{ 0.08 }_{  0.08 }$ &   2.2  &                  FUV-NUV, NUV-g, u-g, g-r, r-i, i-z    \\
   11   &  8  $\pm$ $^{ 143 }_{   5 }$ &  0.52  $\pm$ $^{ 0.74 }_{  0.42 }$ &   1.7  &                       FUV-NUV, NUV-g, g-r, r-i, i-z    \\
   12   & 300  $\pm$ $^{ 101 }_{   1 }$ &  0.44  $\pm$ $^{ 0.06 }_{  0.10 }$ &   1.4  &                  FUV-NUV, NUV-g, u-g, g-r, r-i, i-z    \\
 **15   &  7  $\pm$ $^{ 144 }_{   3 }$ &  0.32  $\pm$ $^{ 0.28 }_{  0.32 }$ &   0.9  &                  FUV-NUV, NUV-g, u-g, g-r, r-i, i-z    \\
 **16   &  8  $\pm$ $^{ 28 }_{   3 }$ &  0.24  $\pm$ $^{ 0.34 }_{  0.24 }$ &   1.2  &                       FUV-NUV, NUV-g, u-g, g-r, r-i    \\
 **17   &  7  $\pm$ $^{ 94 }_{   2 }$ &  0.44  $\pm$ $^{ 0.20 }_{  0.26 }$ &   0.1  &                  FUV-NUV, NUV-g, u-g, g-r, r-i, i-z    \\
 **20   &  5  $\pm$ $^{ 20095 }_{   2 }$ &  0.94  $\pm$ $^{ 0.46 }_{  0.94 }$ &   0.0  &                            FUV-NUV, NUV-g, g-r, r-i    \\
 **21   & 40  $\pm$ $^{ 51 }_{  35 }$ &  0.30  $\pm$ $^{ 0.26 }_{  0.18 }$ &   6.3  &                  FUV-NUV, NUV-g, u-g, g-r, r-i, i-z    \\
   24   & 150  $\pm$ $^{ 51 }_{  101 }$ &  0.22  $\pm$ $^{ 0.30 }_{  0.16 }$ &   1.8  &                  FUV-NUV, NUV-g, u-g, g-r, r-i, i-z    \\
   25   & 2000  $\pm$ $^{ 18100 }_{  1995 }$ &  0.64  $\pm$ $^{ 0.90 }_{  0.48 }$ &   0.6  &                                       g-r, r-i, i-z    \\
 **26   & 40  $\pm$ $^{ 41 }_{  35 }$ &  0.16  $\pm$ $^{ 0.24 }_{  0.14 }$ &   2.4  &                  FUV-NUV, NUV-g, u-g, g-r, r-i, i-z    \\
   27   &  6  $\pm$ $^{ 80 }_{   1 }$ &  0.78  $\pm$ $^{ 0.14 }_{  0.32 }$ &   3.4  &                       FUV-NUV, NUV-g, g-r, r-i, i-z    \\
***29   & 80  $\pm$ $^{ 21 }_{  36 }$ &  0.26  $\pm$ $^{ 0.18 }_{  0.12 }$ &   0.7  &                  FUV-NUV, NUV-g, u-g, g-r, r-i, i-z    \\
\enddata
\tablenotetext{*}{Along the tidal arm/ring.}
\tablenotetext{**}{At the end of the tidal feature.}
\end{deluxetable}

\end{document}